\documentclass[openacc]{rsproca_new}

\usepackage{siunitx}
\usepackage{graphicx}
\usepackage{caption}
\usepackage{subcaption}
\usepackage[dvipsnames]{xcolor}
\usepackage{wrapfig}
\usepackage{comment}
\usepackage{float}
\usepackage{amsmath}
\usepackage{booktabs}
\usepackage{makecell}
\usepackage{adjustbox}
\usepackage{array,makecell}   
\newcolumntype{M}[1]{>{\centering\arraybackslash}m{#1}}

\usepackage{placeins}
\usepackage{nicefrac}
\usepackage{calc}     

\jname{}
\Journal{ }

\begin{document}

\title{Transforming tubular packings to bicontinuous surfaces}

\author{
Vira Raichenko$^{1}$, Alicja Bukat$^{2}$, Michał Bykowski$^{2}$, Łucja Kowalewska$^{2}$, Myfanwy E. Evans$^{1}$}

\address{$^{1}$
Institut für Mathematik, Universität Potsdam\\
14476 Potsdam, Germany\\
Email Address: evans@uni-potsdam.de\\
$^{2}$Faculty of Biology, Department of Plant Anatomy and Cytology,\\ University of Warsaw, Warsaw 02-096, Poland\\}

\subject{Biomaterials, Biomathematics}

\keywords{Minimal surfaces, membrane, cubic rod packings, PLB formation}

\corres{Myfanwy E. Evans\\
\email{evans@uni-potsdam.de}}

\begin{abstract}
  The link between bicontinuous architectures in biological membranes and triply-periodic minimal surfaces (TPMS) is a well established example of stunning geometric form in nature. The prolamellar body (PLB) in early plant plastid development is a classic example, forming the Diamond TPMS in a lipid-protein-pigment membrane. However, the early development of such spectacular geometric structures is poorly understood. Inspired by the presence of tubules in the micrographs of early plastid membrane formation, we explore here geometric modelling of transformations of packings of cylinders that coalesce together to form bicontinuous structures. Using computational modelling, we find that specific cylinder packings with cubic symmetry transform into highly symmetric TPMS, which now stand as a candidate set of surfaces for further investigation into the PLB, as well as other occurrences of bicontinuous membranes.
  


\end{abstract}


\begin{fmtext}

\section{Introduction}
Intricate and complex shapes appear at every scale in biological systems, from molecular assemblies to whole organs. The language of geometry \cite{hyde1996language} enables exploration of a plethora of structures, their formation and their interactions with the environment. Biological membranes are a classic example of the intersection of biology and geometry, where membranes form into highly symmetrical bicontinuous structures that can be modelled by triply periodic surfaces \cite{schoen1970infinite}. For example, the prolamellar body (PLB) in plant plastids takes the form of a Diamond triply-periodic minimal surface (TPMS) (Figure \ref{fig:diamond_PLB}), first described by the mathematician Schwarz in the 1870s \cite{schwarz1871bestimmung} and identified by biologists in 1998 using scattering methods \cite{williams1998x}.


TPMSs are a class of surfaces characterized by zero mean curvature at every point and by periodic repetition along three linearly independent spatial directions. Prominent examples include the Diamond (D) surface and the Gyroid (G) surface, the latter originally described by Alan Schoen \cite{schoen1970infinite}. These surfaces divide space into two continuous, non-intersecting labyrinthine domains, which gives rise to their classification as bicontinuous structures. Among the various TPMS, the Diamond and Gyroid surfaces are especially common in natural systems, where their geometric and topological features support the formation of complex biological and soft-matter structures \cite{jessop2024composite, hyde1984cubic, michielsen2008gyroid}.

\end{fmtext}

\maketitle

\newpage
\noindent

\begin{figure}[H]        
\setlength{\intextsep}{0pt}       
  \setlength{\abovecaptionskip}{4pt} 
  \setlength{\belowcaptionskip}{0pt} 
   \makebox[0pt][l]{
    \begin{minipage}{\textwidth}
    \centering
    \begin{subfigure}[b]{0.3\linewidth}
     \includegraphics[width=1.1\linewidth]{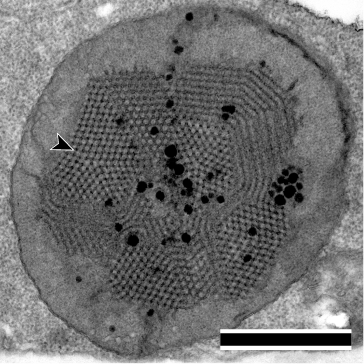}   
      \label{fig:plb_like_d}
    \end{subfigure}\hspace{0.05\linewidth}%
    \begin{subfigure}[b]{0.3\linewidth}
      \includegraphics[width=1.2\linewidth]{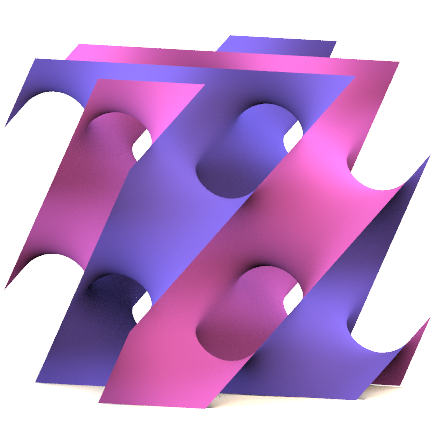}     
      \label{fig:diamond}
    \end{subfigure}
    \caption{(Left) Transmission electron microscopy (TEM) micrograph of an etioplast with a cross-section of the prolamellar body (PLB), indicated by black arrow with white outline. Scale bar = 1~\si{\micro\meter}  (sample fixed according to \cite{bykowski2020spatial}). (Right) The Diamond triply periodic minimal surface.}
    \label{fig:diamond_PLB}
  \end{minipage}}
\end{figure}


Biological interfaces are frequently governed by surface tension, which drives them toward configurations that minimise interfacial area. As a result, many such systems adopt morphologies consistent with area-minimizing surfaces. A substantial body of literature has examined the occurrence of bicontinuous membranes, often referred to as cubic membranes (CMs), in biological literature, and their possible physiological relevance, particularly in cases where these structures correspond to minimal surfaces \cite{almsherqi2009cubic}. Recent advances have elucidated a direct relationship between membrane topology and the mechanical characteristics of lipid bilayers \cite{ridolfi2022nanoscale}, emphasizing the role of geometric constraints in modulating membrane behavior. Notably, TPMS, including Diamond and Primitive (P) surfaces, have been used as reference geometries in the characterization of three-dimensional bicontinuous membranes observed in the mitochondria of Chaos carolinensis \cite{deng1998three}. Additional examples of biologically relevant bicontinuous membranes have similarly been classified as TPMS, underscoring their functional and structural significance across a range of cellular contexts \cite{almsherqi2006cubic}.


Tubular arrangements are structures commonly observed in organelles where cubic membranes form \cite{almsherqi2009cubic}. Imaging and biophysical studies suggest that tubular networks serve as the initial templates for membrane remodelling, though direct observations of tubular-to-cubic transformations have yet to be documented. Steric crowding of proteins on lipid bilayers can induce spontaneous tubulation at length scales set by protein density and membrane tension \cite{stachowiak2010steric}, and ER-resident morphogens such as reticulons and DP1/Yop1p preferentially assemble on tubules to stabilise these curved intermediates \cite{voeltz2006class}. Similarly, in vitro studies observed that a mixture of key PLB lipids and the main protein, light-dependent protochlorophyllide oxidoreductase, in complex with protochlorophyllide and NADPH, results in the spontaneous formation of filaments composed of high-curvature tubules \cite{nguyen2021photocatalytic}. This indicates a direct connection between tubular and cubic arrangements that PLB-building molecules could form. From a geometrical perspective, tubular or cylinderical geometries have some established relationships to high symmetry TPMS. For example, particular packings of cylinders can be completely enveloped by TPMS structures \cite{von1991three}, where the cylinders lie completely within a single labyrinth of the surface.


We will use the presence of tubular structures as precusors to bicontinuous membranes as inspiration to explore the geometric formation of TPMS through cylindrical packings. Our geometric starting point are triply periodic cubic rod packings \cite{o2001invariant}, which have been well documented in structural chemistry and are known to have a close relation to TPMS \cite{evans2013periodic}. We examine infinite, equal-radius cylinders (rods) whose positions remain invariant under the symmetry operations of a cubic space group. We perform computational experiments where we simulate the coalescence of these cylinder packings into periodic minimal surfaces, eventually describing and analysing the resulting TPMS. We use these resulting TPMS as a foundation for comparison to images of PLB membranes formed during plastid development obtained via transmission electron microscopy (TEM). 

\section{Methods}

\begin{figure}[htbp]
  \centering
  \captionsetup{skip=4pt}
  \captionsetup[subfigure]{aboveskip=2pt, belowskip=2pt}

  \begin{subfigure}[b]{0.17\textwidth}
    \includegraphics[width=\linewidth]{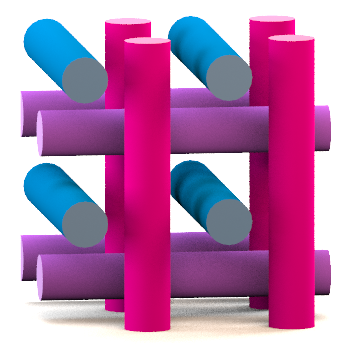}
    \caption{$\Pi^{*}$}
    \label{fig:pi_star_rods}
  \end{subfigure}\hfill
  \begin{subfigure}[b]{0.17\textwidth}
    \includegraphics[width=\linewidth]{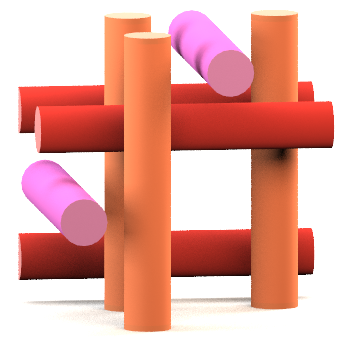}
    \caption{$\Pi^{+}$}
    \label{fig:pi_plus_rods}
  \end{subfigure}\hfill
  \begin{subfigure}[b]{0.17\textwidth}
    \includegraphics[width=\linewidth]{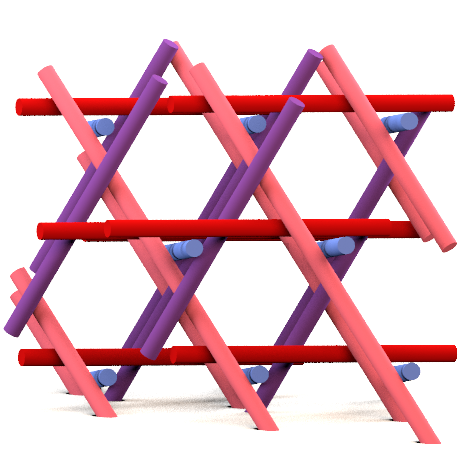}
    \caption{$\Sigma^{+}$}
    \label{fig:sigma_rods}
  \end{subfigure}\hfill
  \begin{subfigure}[b]{0.17\textwidth}
    \includegraphics[width=\linewidth]{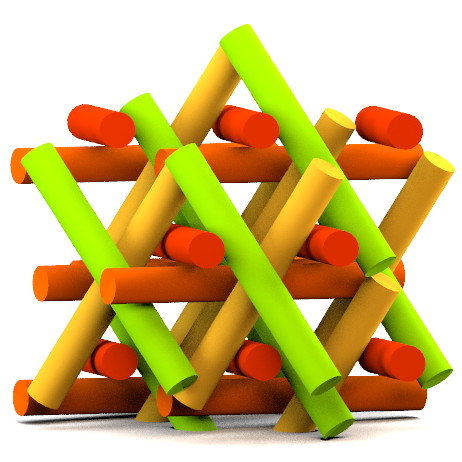}
    \caption{$\Omega^{+}$}
    \label{fig:omega_rods}
  \end{subfigure}\hfill
  \begin{subfigure}[b]{0.17\textwidth}
    \includegraphics[width=\linewidth]{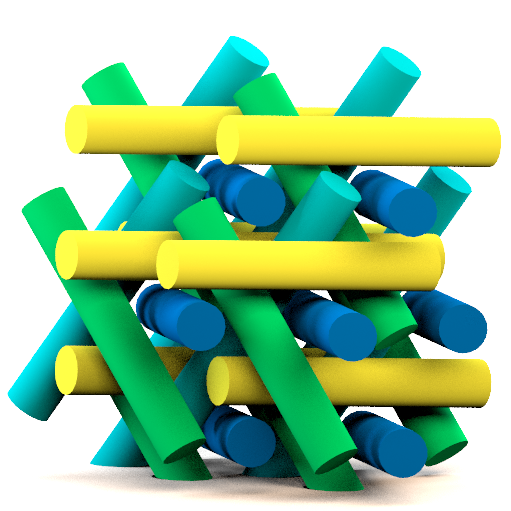}
    \caption{$\Gamma$}
    \label{fig:gamma_rods}
  \end{subfigure}

  \caption[Compact rod‐packing configurations]{The invariant cubic rod‐packing. Rod axes along  $\langle100\rangle$: (a)~$\Pi^{*}$ and (b)~$\Pi^{+}$. Rod axes along  $\langle111\rangle$: (c)~$\Sigma^{+}$, (d)~$\Omega^{+}$, and (e)~ $\Gamma$.}
  \label{fig:rod_pack_combined}
\end{figure}

We examine infinite, equal-radius cylinders (rods) whose positions remain invariant under the symmetry operations of a cubic space group. These include five invariant cubic packings $\Pi^{*}$, $\Pi^{+}$, $\Sigma^{+}$, $\Omega^{+}$, and $\Gamma$ shown in Figure~\ref{fig:rod_pack_combined} \cite{o2001invariant}. Their rod axes run either along the $\langle100\rangle$ or the $\langle111\rangle$ directions of the cubic unit cell, and their fractional coordinates are listed below.

For each rod packing, we construct a mesh of a single unit cell (Table~\ref{Tab:Table_surfaces}).  We build the rods using formulas provided by \cite{o2001invariant}. The rods are inflated until they touch; the resulting intersections are removed, forming a surface that divides space into interior-rod and exterior regions. This resulting surface is optimised in periodic boundary conditions using \textsc{Surface Evolver}~\cite{brakke1994surface} following the robust scheme of Himmelmann \cite{himmelmann2025amorphous}; we minimise the Willmore energy, initially including a Gaussian curvature term while progressively refining the mesh to achieve smooth, stable convergence. The functional used for this optimisation process of a mesh~$M$ is given by:
\begin{equation}
  E(M) \;=\; \int_M H^2\,\mathrm dA \;+\; \alpha(n)\int_M K^2\,\mathrm dA,
\end{equation}
where \(H\) and \(K\) are the mean and Gaussian curvatures, respectively. The initial term in the energy functional is the Willmore energy, which is commonly employed in the study of minimal surfaces due to its sensitivity to mean curvature. However, it has a tendency to constrict neck-like regions in tubular geometries. To mitigate this effect during early stages of the optimisation, we incorporate a Gaussian curvature term, which helps maintain a more uniform channel structure. Once a balanced morphology is established, this term is removed and the optimisation proceeds using only the Willmore energy until changes in the total energy become negligible.
Additionally, the minimisation process begins with a volume constraint to ensure equal partitioning of the two labyrinthine domains. This constraint is lifted after a stable and symmetric mesh is achieved. This two-stage approach yields well-behaved minimisers that preserve both geometric balance and numerical stability.

Topology analysis of TPMS is performed by analysing two interpenetrating networks – its ‘labyrinth graphs’. Each labyrinth is first isolated as a single connected component in the voxelised image.  
The skeletonisation is performed with MATLAB’s \texttt{bwskel} function, which implements the medial-axis transform~\cite{lee1994building}. We use skeletonisation as a way to approximate the medial axis topology via a 1-dimensional structure for simplicity and robustness.
The voxel skeleton is converted into vertices and edges via each vertex at a voxel centre, neighbouring voxels joined by edges, nd then smoothed with ten iterations of Laplacian smoothing~\cite{sorkine2004laplacian}.  
Because the approximation of medial axis can be unstable, we collapse any segment that contains fewer than eight vertices between branch points. The final obtained graph is then analysed by computing the degrees of the vertices, i.e., the coordination numbers of the net, detecting minimal cycle lengths, and inferring their space symmetry groups.

\section{Results}

For each rod packing shown in Figure~\ref{fig:rod_pack_combined}, we cut out a cubic unit cell (Table~\ref{Tab:Table_surfaces}).  
Minimising the Willmore energy then drives the mesh toward a minimal surface.  
During the optimization we do not consolidate any points, edges, or faces, which allows \textsc{Surface Evolver}  to preserve the original genus of the surface, and the resulting surfaces retain the same periodic boundary conditions. Each initial surface inherits the topology of the underlying linked rods \cite{rosi2005rod}. In other words, the net is the initial rod‐packing, with extra edges running through the necks where the rod surfaces come into contact. We will sometimes call these structures linked rod packings. After the evolution, we obtain minimal surfaces, and Table~\ref{Tab:Table_surfaces} lists the corresponding identifications. The files for the final surfaces are provided in the supplementary materials.

\begin{table}[htbp]
  \centering
  \renewcommand{\arraystretch}{1.6}  
  \setlength{\tabcolsep}{4pt}        
  \begin{tabular}{l  c  c  c  l}
    \hline
     & Initial surface
     & Medial axis / Space group
     & Evolved surface
     & Surface name/G–H \\
    \hline
    $\Pi^{*}$   
      & \includegraphics[height=2.2cm]{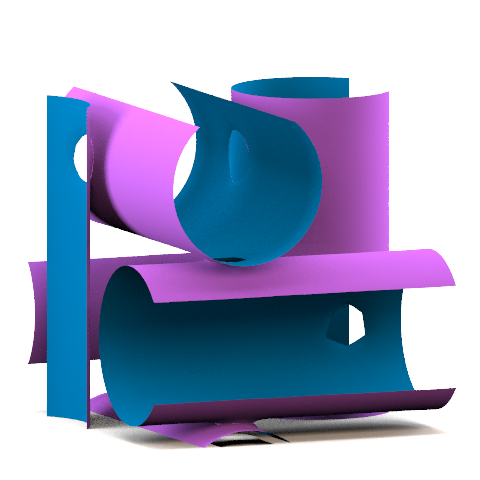}
      & \makecell{\textit{nbo}/ \textit{bcu} \\[-3pt]\small $Im\bar{3}m$}
      & \includegraphics[height=2.2cm]{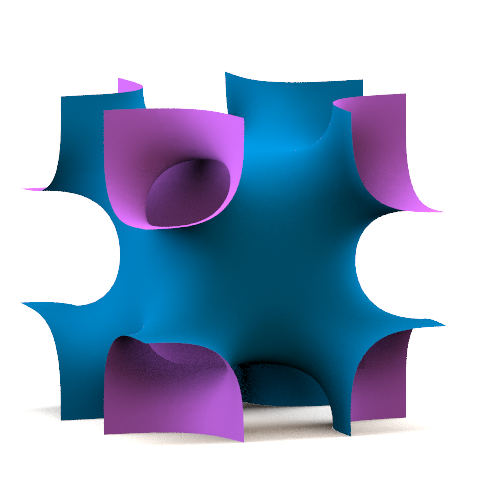}
      & \makecell{I-WP\\[-3pt]\small $Im\bar{3}m\text{–}Im\bar{3}m$}  \\ 
    \hline
    $\Pi^{+}$   
      & \includegraphics[height=2.2cm]{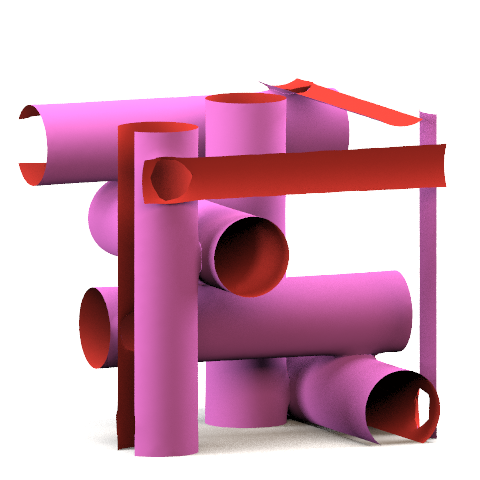}
      & \makecell{\textit{bmn} / –\\[-3pt]\small $I4_{1}32$}
      & \includegraphics[height=2.2cm]{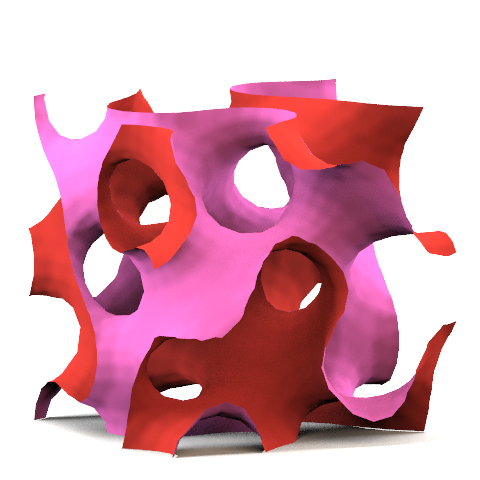}
      & \makecell{$C(^\pm Y)$\\[-3pt]\small $Ia\bar{3}\text{–}Pa\bar{3}$}  \\ 
    \hline
    $\Sigma^{+}$
      & \includegraphics[height=2.2cm]{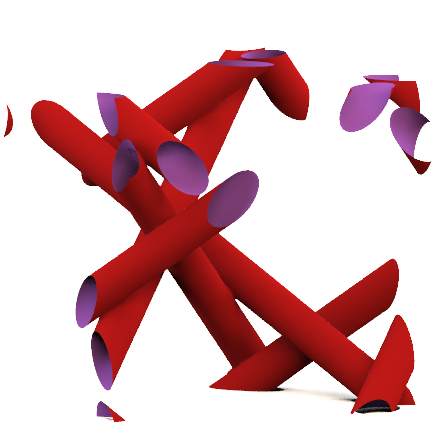}
      & \makecell{\textit{sgn} / –\\[-3pt]\small $I4_{1}32$}
      & \includegraphics[height=2.2cm]{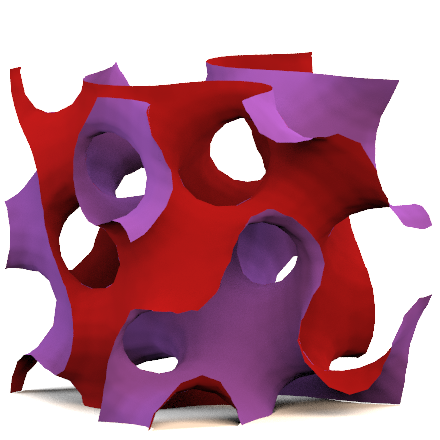}
      & \makecell{$C(^\pm Y)$\\[-3pt]\small $Ia\bar{3}\text{–}Pa\bar{3}$}  \\ 
    \hline
    $\Omega^{+}$
      & \includegraphics[height=2.2cm]{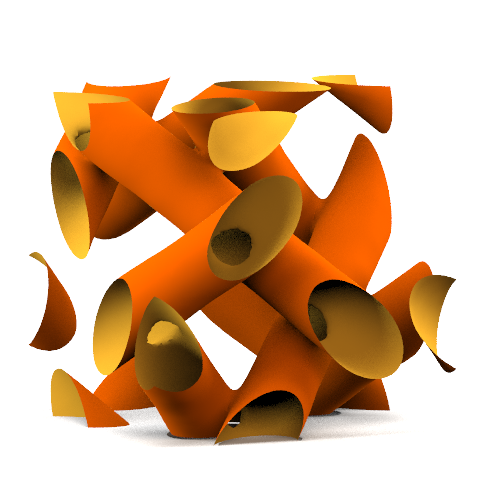}
      & \makecell{\textit{utb-z} / –\\[-3pt]\small $I432$}
      & \includegraphics[height=2.2cm]{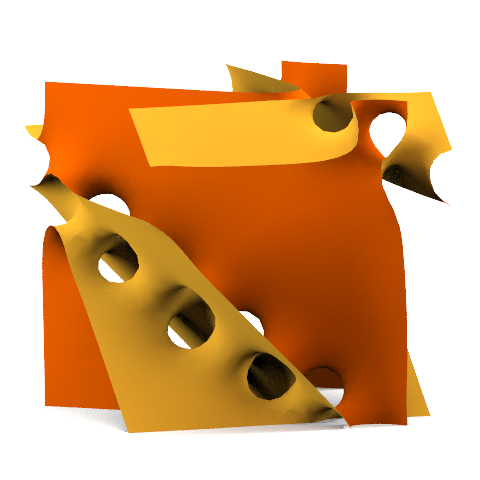}
      & \makecell{H\\[-3pt]\small $P\bar{6}m2\text{–}P\bar{6}m2$}  \\ 
    \hline
    $\Gamma$    
      & \includegraphics[height=2.2cm]{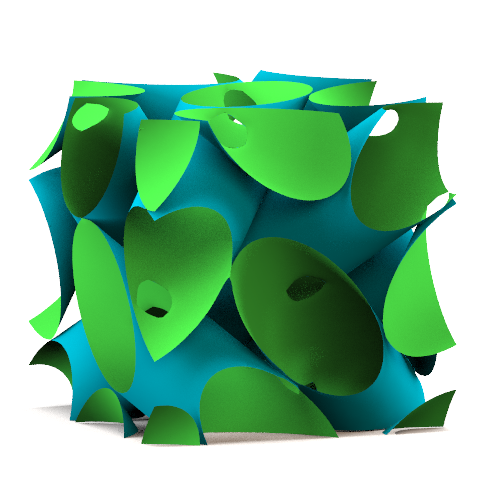}
      & \makecell{\textit{gan} / \textit{lcs}\\[-3pt]\small $Ia\bar{3}d$}
      & \includegraphics[height=2.2cm]{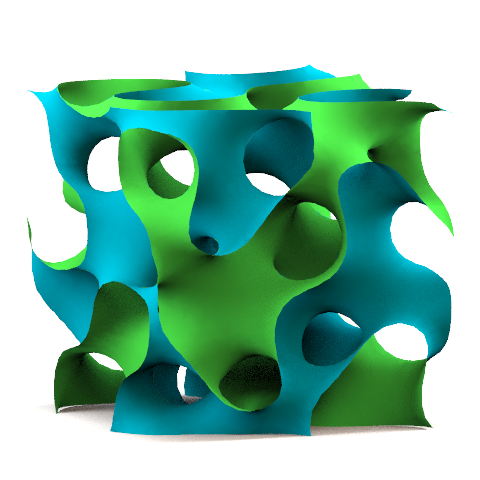}
      & \makecell{$C(I_{2}\!-\!Y^{**})$\\[-3pt]\small $Ia\bar{3}d\text{–}Ia\bar{3}d$}  \\ 
    \hline
  \end{tabular}
   \caption{%
The summary of the five rod–packing geometries to minimal surfaces transitions. For each packing (first column), the table shows: 
(a) the \emph{initial surface} generated directly from the rod–packing skeleton, 
(b) the corresponding \emph{rod net} and corresponding dual net for each initial surface and their crystallographic \emph{space group}, 
(c) the \emph{evolved minimal surface}, and 
(d) The identified \emph{minimal‐surface}: $G$ denotes the symmetry group of the minimal surface, and $H$ denotes the symmetry group of its labyrinth network.
}

  \label{Tab:Table_surfaces}
\end{table}

We compare the resulting minimal surfaces with known examples by constructing the approximation of medial axis of each labyrinth.  
Because the medial axis is highly sensitive to small topological variations, we collapse any edge chain shorter than eight vertices before analysing the underlying graph.  
The resulting graphs are then matched to reference nets in the RCSR database~\cite{o2008reticular}.

\subsubsection{$\Pi^*$}
The rod packing $\Pi^*$ consists of rods running along $\langle100\rangle$ with symmetry $Pm\bar{3}n$ (Figure \ref{fig:pi_star_rods}): $(\tfrac12,0,u),;(u,\tfrac12,0),;(0,u,\tfrac12)$. When we inflated the rods and pinched the surface at the contact points, we obtained a surface that separates the space into two connected components with symmetry $Im\bar{3}m$. The region enclosed by this surface can be described by the \textit{nbo} net~\cite{o2008reticular}, which traces along its skeleton and describes its topology. The complementary region outside of the surface can be described by the \textit{bcu} net~\cite{o2008reticular}.

\begin{figure}[htbp]
  \centering
  \setlength{\abovecaptionskip}{4pt}
  \setlength{\belowcaptionskip}{4pt}
  \vspace{-4pt}

  \begin{subfigure}[b]{0.25\linewidth}
    \centering
    \includegraphics[width=\linewidth]{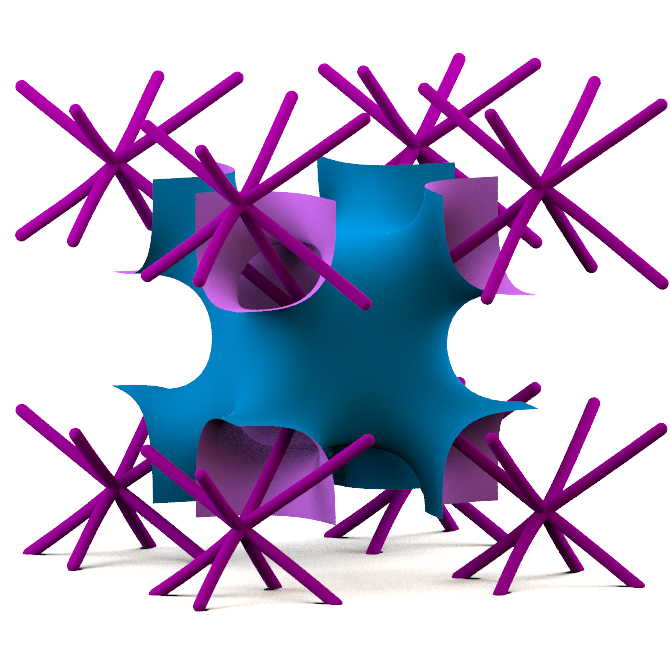}
    \caption{}
    \label{fig:pi_star_surf}
  \end{subfigure}\hspace{0.05\linewidth}%
  \begin{subfigure}[b]{0.25\linewidth}
    \centering
    \includegraphics[width=0.9\linewidth]{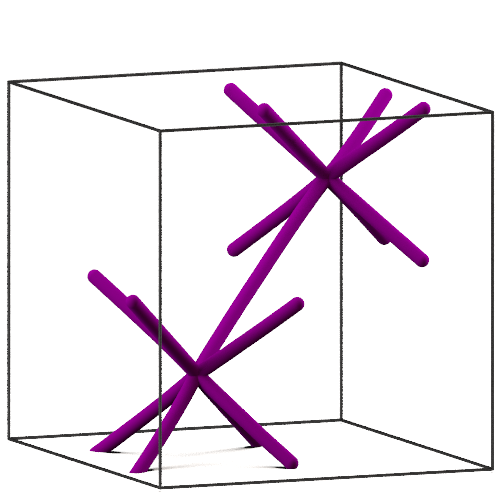}
    \caption{}
    \label{fig:iwp}
  \end{subfigure}
    
  \vspace{-4pt}
  \caption{Evolution of a $\Pi^{*}$ rod packing into the I–WP minimal surface and its labyrinth graph. (a) The relaxed I–WP surface obtained from the $\Pi^{*}$ cubic unit cell as an input. (b) The corresponding \textit{bcu}  net extracted from the surface’s medial axis, illustrating the labyrinth connectivity.}

  \label{fig:pi_star}
\end{figure}

After minimisation, the topology and geometry of the channels of the minimal surface do not change siginificantly enough to alter the network representations of the channels: they remain as the \textit{nbo} and \textit{bcu} nets.  
Consequently, the minimal surface is readily identifiable as the well-known I-WP minimal surface~\cite{schoen1970infinite}. The resulting surface along with the \textit{bcu} net is shown in Figure \ref{fig:pi_star}.

\subsubsection{$\Pi^+$}
The second case of a rod packing where rods run along $\langle100\rangle$ is $\Pi^+$ with symmetry $I4_{1}32$: $\Pi^{+}: \; (\tfrac14,0,u),\allowbreak\;(\tfrac34,\tfrac12,u),\allowbreak\;(u,\tfrac12,0),\allowbreak\;(u,\tfrac34,\tfrac12),\allowbreak\;(0,u,\tfrac14),\allowbreak\;(\tfrac12,u,\tfrac12)$.
 In this case, the inflated tubular surface has the same symmetry group as the underlying rod packing and the channel inside the tubes exhibits the connectivity of the \textit{bmn} net~\cite{o2008reticular}.

\begin{figure}[htbp]
  \centering
  \setlength{\abovecaptionskip}{4pt}
  \setlength{\belowcaptionskip}{4pt}

  \begin{subfigure}[b]{0.25\linewidth}
    \centering
    \includegraphics[width=\linewidth]{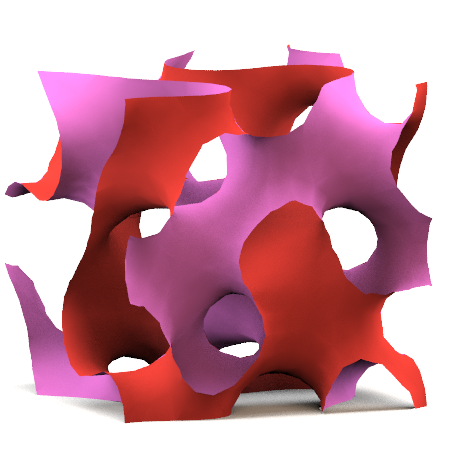}
    \caption{}
    \label{fig:pi_plus_surf}
  \end{subfigure}\hspace{0.01\linewidth}%
  \begin{subfigure}[b]{0.25\linewidth}
    \centering
    \includegraphics[width=\linewidth]{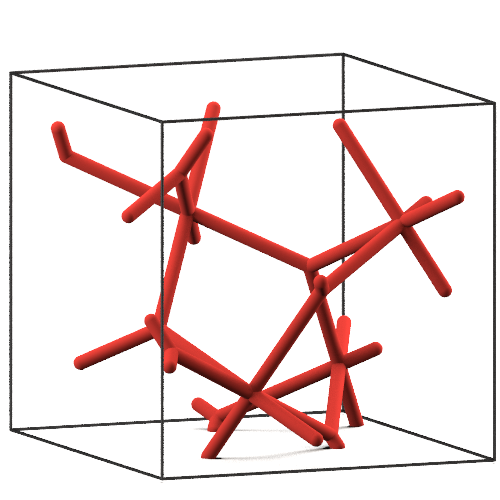}
    \caption{}
    \label{fig:pyr}
  \end{subfigure}\hspace{0.01\linewidth}%
  \begin{subfigure}[b]{0.25\linewidth}
    \centering
    \includegraphics[width=\linewidth]{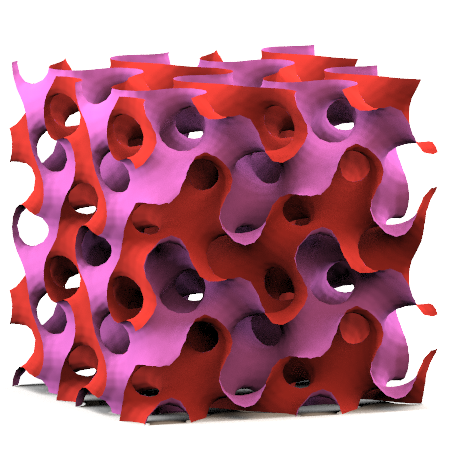}
    \caption{}
    \label{fig:cy}
  \end{subfigure}

  \caption{(a) The evolved surface obtained from a cubic unit cell enveloping the linked $\Pi^{+}$ rod packing. (b) The self‐dual \textit{pyr} network corresponding to the resulting surface. (c) Two unit cells of the $C(^\pm Y)$ minimal surface, which coincide with the surface in (a).}
  \label{fig:pi_plus}
\end{figure}

After minimisation, the skeleton of either of a channel labyrinth of the resulting minimal surface (Figure \ref{fig:pi_plus_surf}) contains vertices of degrees~3 and~6, described by the self-dual net \textit{pyr}~\cite{o2008reticular} with symmetry  $Pa\bar{3}$. Having the \textit{pyr} net as the topology of the channels indicates that the resulting surface is the $C(^{\pm}Y)$ minimal surface~\cite{lord2003periodic} with $Ia\bar{3}$ symmetry, as shown in Figure \ref{fig:pi_plus}.


\subsubsection{$\Sigma^+$}

The rod packing $\Sigma^+$ has rod axes in the $\langle111\rangle$ directions, with symmetry $I4_{1}32$:
$ (\tfrac13+u,\tfrac23+u,u),\allowbreak\;(\tfrac23-u,\tfrac56-u,u),\allowbreak\;(\tfrac16+u,\tfrac23-u,u),\allowbreak\;(\tfrac56-u,\tfrac56+u,u)$. The surface that comes from the inflation and coalescence of the rods preserves the symmetry group from the $\Sigma^+$ packing -- $I4_{1}32$. The net describing the connectivity of the space enclosed by the rod surface is the \textit{sgn} net~\cite{o2008reticular}. During the smoothing and optimisation process, this tubular surface converges to the $C(^{\pm}Y)$ minimal surface, the same surface as was formed by the $\Pi^+$ packing, as shown in Figure \ref{fig:pi_plus}. The close relationship between \(\Pi^{+}\) and \(\Sigma^{+}\) has already been previously noted~\cite{hyde2008short}.

\begin{figure}[htbp]
  \centering
  \setlength{\abovecaptionskip}{4pt}
  \setlength{\belowcaptionskip}{4pt}

  \begin{subfigure}[b]{0.23\linewidth}
    \centering
    \includegraphics[width=\linewidth]{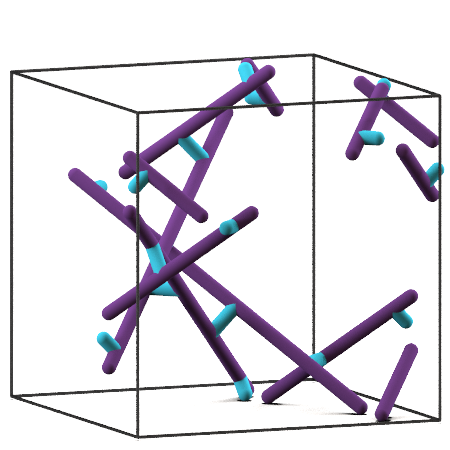}
    \caption{}
    \label{fig:pi_plus_surf}
  \end{subfigure}\hspace{0.02\linewidth}%
  \begin{subfigure}[b]{0.23\linewidth}
    \centering
    \includegraphics[width=\linewidth]{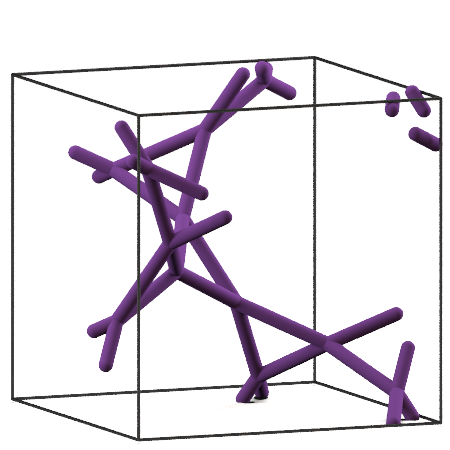}
    \caption{}
    \label{fig:lidi_med}
  \end{subfigure}\hspace{0.02\linewidth}%
  \begin{subfigure}[b]{0.23\linewidth}
    \centering
    \includegraphics[width=\linewidth]{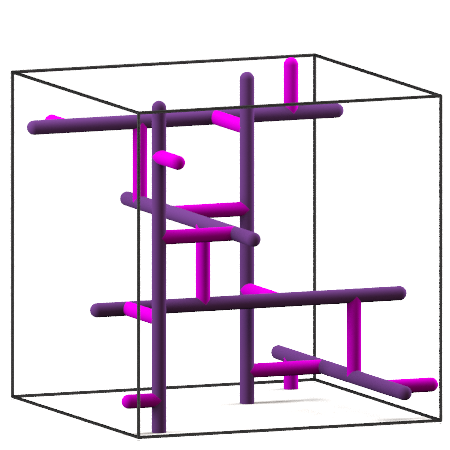}
    \caption{}
    \label{fig:lidinoid}
  \end{subfigure}

  \caption{Deformation from (a) \textit{sgn} via (b) \textit{lcv} to (c) \textit{bmn}: the blue edges in the \textit{sgn} network are first contracted to points and then re‐expanded via the pink edges to form the \textit{bmn} network.}
  \label{fig:sgn-bmn}
\end{figure}

The common surface of the inflated \(\Pi^{+}\) and \(\Sigma^{+}\) packings can be demonstrated through a related network to the channel connectivity of the rod surfaces of both structures. Collapsing each rod–rod edge of the \textit{sgn} net to their midpoints converts the \textit{sgn} net to the \textit{lcv} net~\cite{o2008reticular}, which can subsequently be converted into the \textit{bmn} net though vertex splitting the degree-4 vertices. The progression is shown in Figure~\ref{fig:sgn-bmn}.

\subsubsection{$\Omega^+$}
Another rod packing with rods in the $\langle111\rangle$ directions is denoted as $\Omega^+$ and has $I432$ symmetry. It can be described by the following rod axes in space: $ (\tfrac13+u,\tfrac23+u,u),\;(\tfrac23-u,\tfrac13-u,u),\;(\tfrac23+u,\tfrac23-u,u),\;(\tfrac13-u,\tfrac13+u,u)$. This rod packing can be inflated and connected, resulting in a region of space enclosed by the tubular surface, which has connectivity of the \textit{utb–z} net~\cite{o2008reticular}, while the space outside the rods follows the \textit{sod} net~\cite{o2008reticular}.

\begin{figure}[htbp]
  \centering
  \begin{subfigure}[b]{0.22\textwidth}
    \centering
    \includegraphics[width=\linewidth]{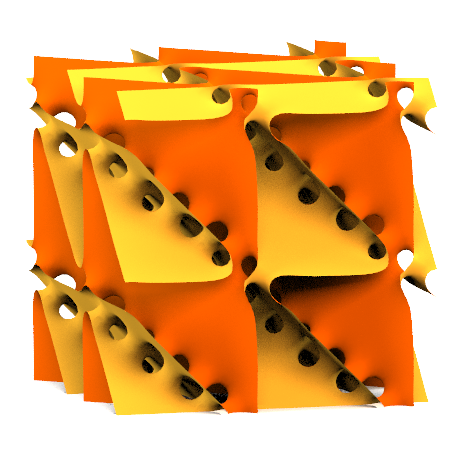}
    \caption{}
    \label{fig:omega_surf}
  \end{subfigure} 
  \begin{subfigure}[b]{0.22\textwidth}
    \centering
    \includegraphics[width=\linewidth]{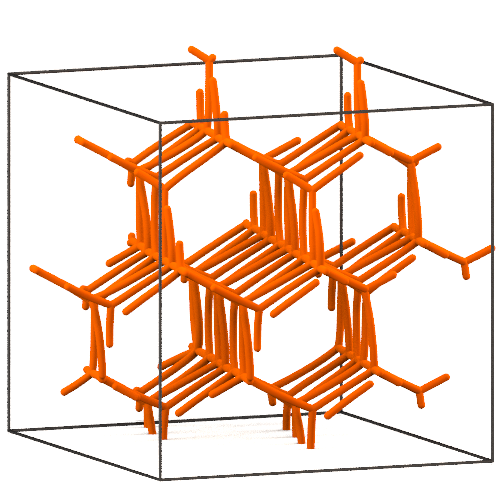}
    \caption{}
    \label{fig:med_omega}
  \end{subfigure}
  \begin{subfigure}[b]{0.22\textwidth}
    \centering
    \includegraphics[width=\linewidth]{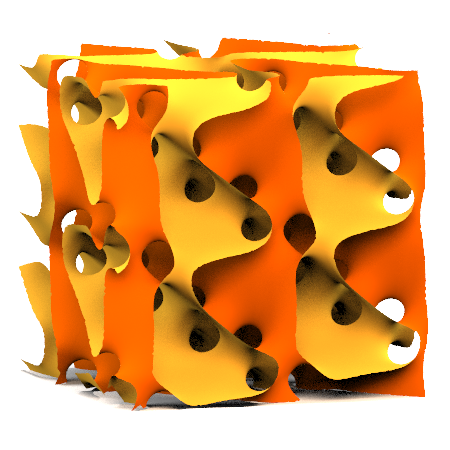}
    \caption{}
    \label{fig:h_surface}
  \end{subfigure}
  \begin{subfigure}[b]{0.22\textwidth}
    \centering
    \includegraphics[width=\linewidth]{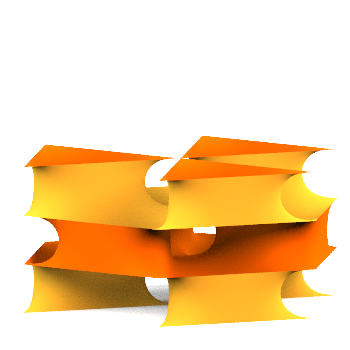}
    \caption{}
    \label{fig:h_small}
  \end{subfigure}
  \caption{Demonstration that $\Omega^+$ evolves into H‐surfaces. (a) Surface obtained by relaxing the $\Omega^{+}$ rod packing, identified as an H‐surface. (b) The hms net describing the topology of both labyrinths of the H‐surface. (c) The H‐surface with a more balanced choice of the free parameter that stretches the surface, oriented to match the view in (a). (d) The same H‐surface is oriented in the more conventional presentation. }
  \label{fig:omega}
\end{figure}

On optimisation, the tubular surface converges to a minimal surface. To confirm its identity, we extracted the approximation of medial axes of the two interwoven labyrinths. The result is the \textit{hms} net in each channel (Figure~\ref{fig:med_omega}); this minimal net \cite{delgado2013minimal} corresponds to a member of the H-surface family first catalogued by Schwarz in 1867 \cite{schwarz1871bestimmung}. Figure~\ref{fig:omega_surf} shows a patch built from two unit cells of the $\Omega^{+}$ rod packing. Although the H-family has no single canonical form, our reconstruction comes remarkably close to the six-ended Scherk (saddle-tower) limit \cite{karcher1988embedded}.


In this case, the specific evolution of the surface under optimisation is also itneresting. While the final structure is independent of the choice of unit cell, it is sensitive to the energy functional; the surface first drifts toward the O,C-TO Schoen surface, which is an intuitive outcome given that the outer channel’s \textit{sod} topology is known to realise that geometry. Yet prolonged Willmore-energy minimisation drives the system instead to the H-surface.

\subsubsection{$\Gamma$}

The final rod packing that we consider, $\Gamma$, is the densest out of all with symmetry $Ia\bar{3}d$: $(u,u,u),\allowbreak\;(\bar{u},\tfrac12-u,u),\allowbreak\;(\tfrac12+u,\bar{u},u),\allowbreak\;(\tfrac12-u,\tfrac12+u,u)$. The surface that is obtained by inflating the rods and connecting them has an inner channel described by the \textit{gan} network and the complementary \textit{lcs}, which makes the tubular surface of the same symmetry as the initial rod packing. 

\begin{figure}[htbp]
  \centering
  \setlength{\tabcolsep}{1pt}      
  \renewcommand{\arraystretch}{0}   

  \begin{tabular}{ccc}
    \subcaptionbox{\label{fig:gamma_model}}
      {\includegraphics[width=.25\linewidth]{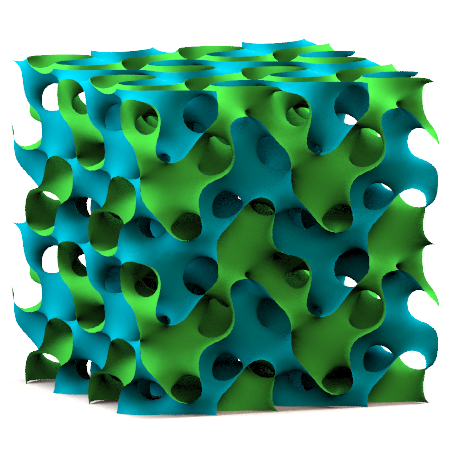}} &
    \subcaptionbox{\label{fig:gan}}
      {\includegraphics[width=.25\linewidth]{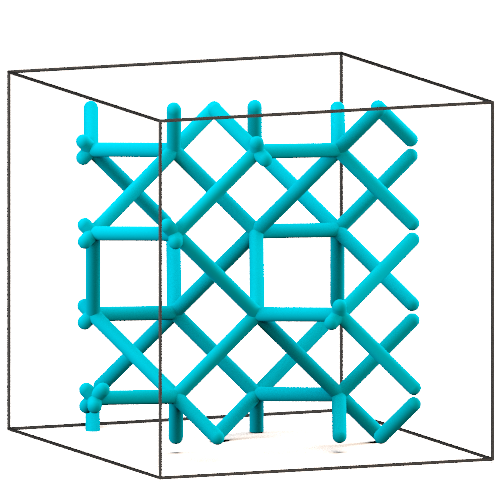}} &
    \subcaptionbox{\label{fig:lcs}}
      {\includegraphics[width=.25\linewidth]{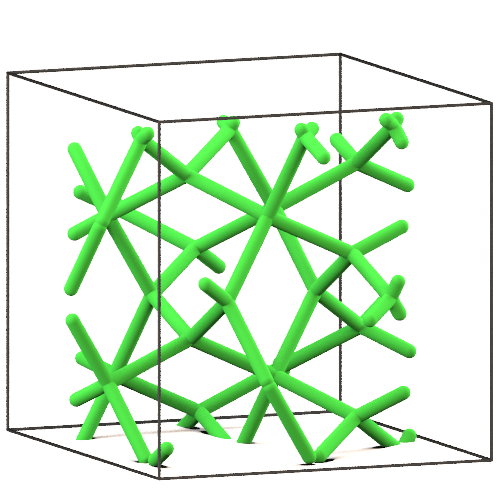}}
  \end{tabular}

  \caption{Topology and intermediate geometry of the $\Gamma$ surface. (a) Simulated minimal surface from the  $\Gamma$ packing : \mbox{$C(I_2\!-\!Y^{**})$}. The unit cells of the dual nets that describe the topology of (a) are (b) \textit{gan} and (c) \textit{lcs}.}
  \label{fig:gamma_topology}
\end{figure}

The surface optimises to a minimal surface while retaining the original labyrinth topology of \textit{gan} and \textit{lcs} (Figure \ref{fig:gamma_model}). Both nets possess $Ia\bar{3}d$ symmetry, as does the minimal surface. This surface was first reported by von Schnerring \cite{von1991three} as a periodic nodal surface that envelops the $\Gamma$ rod packing, called $C(I_2\!-\!Y^{**})$, approximated by the Fourier series

\begin{align}
\label{eg:CI_Y}
C(I_2\!-\!Y^{**})(x,y,z)
&=-2\Bigl[\sin x \cos y \sin z
      +\sin x \,\sin(2y)\,\cos z
      +\cos x \,\sin y \,\sin(2z)\Bigr]\notag\\
&\quad+\cos(2x)\cos(2y)
      +\cos(2y)\cos(2z)
      +\cos(2z)\cos(2x).
\end{align}

We use the $C(I_2\!-\!Y^{**})$ terminology for the periodic nodal surface to also denote the minimal surface here. Topologically, the corresponding minimal surface is a double gyroid connected by extra channels inserted along the $\langle111\rangle$ directions of space to impose inversion symmetry, giving the space group $Ia\bar{3}d$ \cite{wohlgemuth2001triply}. A standard gyroid is threaded by two interpenetrating \textit{srs} nets; here, the \textit{gan} labyrinth comprises two such dual \textit{srs} copies joined by straight lines that pass through their vertices along the directions $[-1,1,-1]$, $[1,1,1]$, $[1,1,-1]$, and $[-1,1,1]$. The complementary labyrinth is described by the \textit{lcs} net. Interestingly, $C(I_2\!-\!Y^{**})$ structure was suggested to be the most favourable for the production of photonic band-gap materials \cite{michielsen2003photonic}.

\section{Biology inspired arrangments}
\label{sec:bio}

Taking the geometric considerations of transforming tubular packings into periodic minimal surfaces, we consider these ideas in the context of membrane formation. The formation pathways of cubic membranes remain largely unknown. These highly ordered membrane structures are typically observed only in their fully developed state. High-resolution imaging using TEM provides only 2D snapshots of 4D reality, limiting our ability to reconstruct and understand the spatial basis of dynamic processes involved in their formation. 
Nevertheless, various precursor membrane arrangements can be detected prior to the emergence of fully developed cubic membranes. 

\begin{figure}[htbp]
  \centering
  \includegraphics[width=0.9\textwidth]{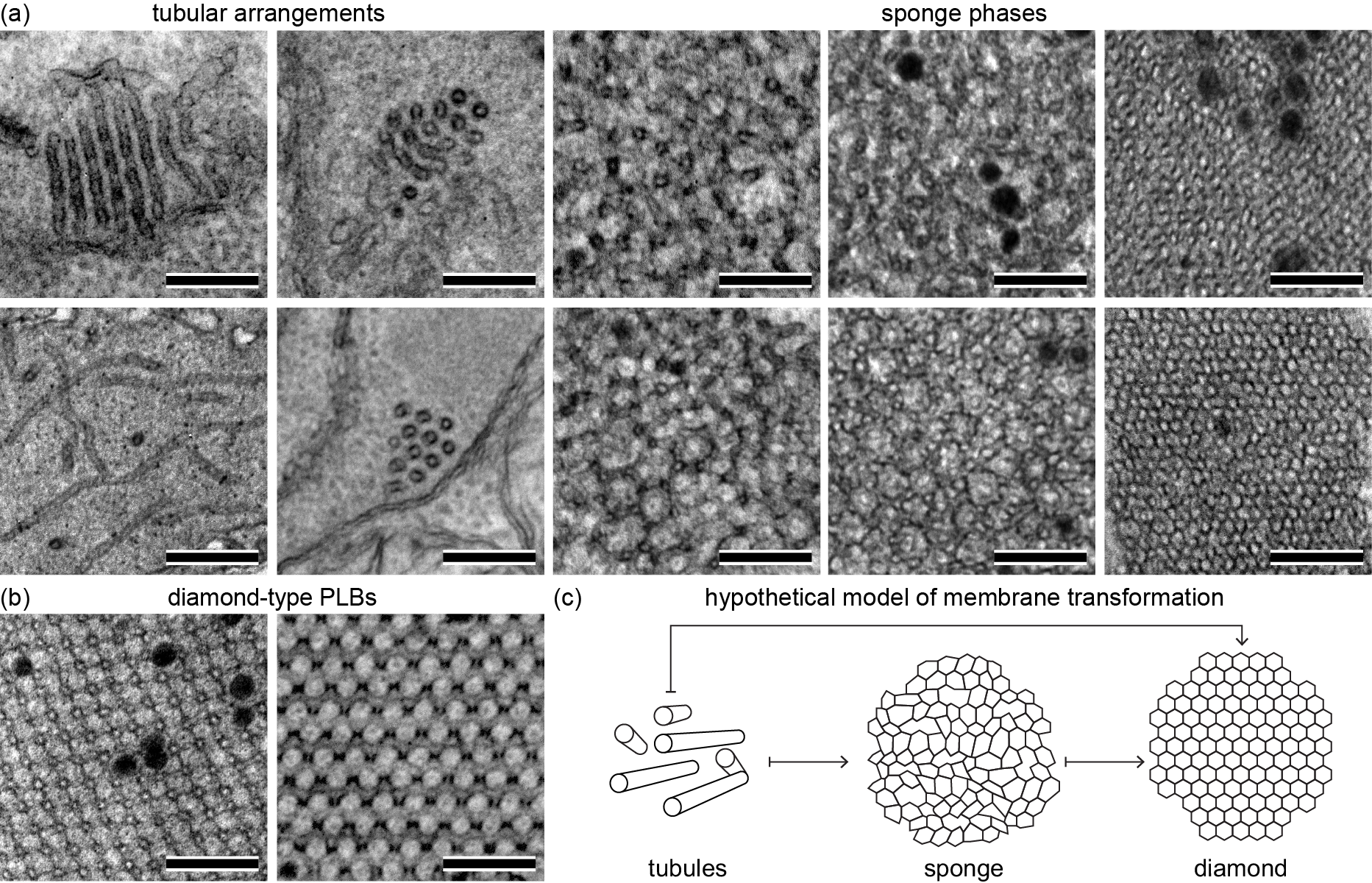}
  \caption{Membrane tubule-to-surface transitions during plastid development in angiosperms.
  (a) Transmission electron microscopy (TEM) micrographs (samples fixed according to \cite{bykowski2020spatial}) showing precursor structures of prolamellar bodies (PLBs) observed during early stages of plastid development. \textit{Left side}: various tubular membrane arrangements—ranging from highly regular to more disordered—observed prior to the appearance of the sponge-like phase and fully developed diamond-type PLBs. \textit{Right side}: representative examples of sponge-like membrane structures exhibiting varying levels of organization and morphological resemblance to the diamond-type PLB, observed before the formation of fully developed PLBs. (b) Fully developed PLBs exhibiting a diamond-type architecture. (c) Schematic representation of the observed transitions, illustrating potential pathways from tubular arrangements to sponge phases, from sponge phases to diamond-type PLBs, or directly from tubules to diamond-type PLBs. Note that the schematic is not drawn to scale. Scale bar = 200 nm (a-b).}
  \label{fig:tubules_figure}
\end{figure}

We focus here specifically on tubular membrane architectures (\ref{fig:tubules_figure}a), which we observed to appear before the formation of the Diamond-type PLBs (\ref{fig:tubules_figure}b). Additionally, we detected complex tubular arrangements (highly regular and disordered) preceding the development of less ordered sponge-like phases, which themselves emerge before the formation of fully developed PLBs (\ref{fig:tubules_figure}a). These sponge phases bear morphological resemblance to the Diamond-type structure but lack crystalline order at this stage.
PLBs serve as a unique example of a naturally occurring cubic membrane based on a Diamond minimal surface, forming within etioplasts—a type of plastid that develops in dark-grown seedlings of plants (\cite{gunning1975ultrastructure}, \cite{gunning2001membrane}, \cite{kowalewska2019spatial}). The PLB exhibits a bicontinuous architecture consistent with a Diamond nanomorphology (\cite{hain2022spire}) and serves as a structural and functional precursor to the photosynthetically active thylakoid membrane system found in mature chloroplasts.
Current analytical tools, such as SPIRE (\cite{hain2022spire}), enable confident identification of such diamond-type arrangements; however, they are not yet applicable for characterizing sponge phases, leaving their precise topology unresolved.
This work addresses a fundamental topological question with direct biological relevance: Can membrane tubules reorganize into a minimal surface, potentially representing an early stage in cubic membrane formation? While our immediate focus is on the development of the PLB, the broader implications extend across biological systems. Cubic membranes are found in all kingdoms of life and are frequently associated with stress conditions or specific stages of development (\cite{almsherqi2009cubic}). Notably, tubular membrane arrangements are commonly observed in cells (\cite{gachie2025vipp1}, \cite{rafelski2013mitochondrial}, \cite{roth2019arbuscular}, \cite{wang2022endoplasmic}, \cite{zhang2015directional}), including those in which cubic membranes have not been conclusively identified. This raises the possibility that tubules may act as universal precursors to cubic membranes (\ref{fig:tubules_figure}c)—both as part of normal organelle development and as structural adaptations to environmental stress.

\begin{figure}[H]
  \centering
  \setlength{\abovecaptionskip}{4pt}
  \setlength{\belowcaptionskip}{4pt}

  \begin{subfigure}[b]{0.25\linewidth}
    \centering
    \includegraphics[width=\linewidth]{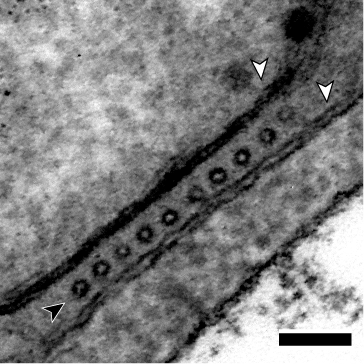}
    \caption{}
    \label{fig:real_data}
  \end{subfigure}\hspace{0.02\linewidth}%
  \begin{subfigure}[b]{0.25\linewidth}
    \centering
    \includegraphics[width=\linewidth]{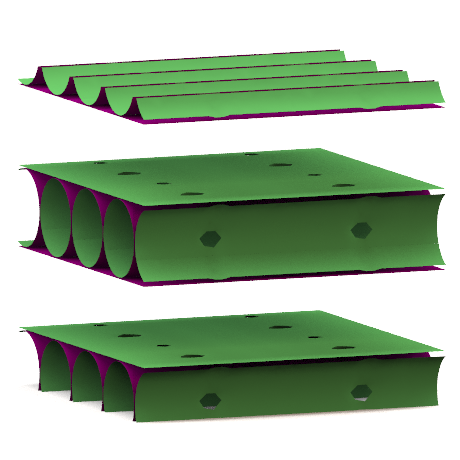}
    \caption{}
    \label{fig:real_model}
  \end{subfigure}\hspace{0.02\linewidth}%
  \begin{subfigure}[b]{0.25\linewidth}
    \centering
    \includegraphics[width=\linewidth]{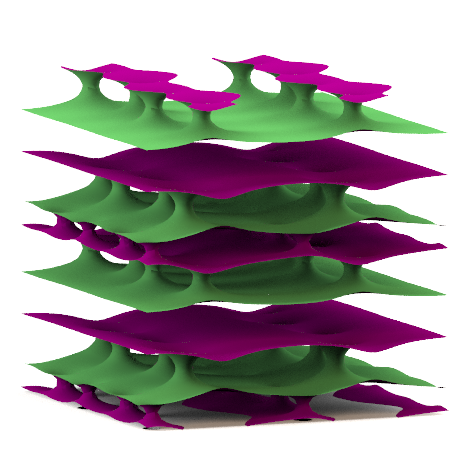}
    \caption{}
    \label{fig:real_formed}
  \end{subfigure}
  \caption{Demonstration of surface simulation from tubes and sheets. (a) Transmission electron microscopy (TEM) micrograph showing a combination of tubules and lamellae in proplastids of \textit{Arabidopsis thaliana} at early developmental stages, preceding prolamellar body (PLB) formation (sample fixed according to \cite{bykowski2020spatial}). Black arrow with white outline indicate tubules, while white arrows with black outlines indicate lamellae. Scale bar = 100 nm. (b) Initial arrangement of cylinders and sheets matching the template geometry. (c) Relaxed minimal surface after energy minimisation.}
  \label{fig:real}
\end{figure}

In addition to purely tubular arrangements, combinations of tubules and lamellae can also occur. Such an arrangement is observed in proplastids of \textit{Arabidopsis thaliana} during early developmental stages. Later, these proplastids develop PLBs, suggesting that the tubular–lamellar architecture may represent a precursor structure to the PLB. Figure \ref{fig:real_data} demonstrates an example of this organization, in which tubules are packed between lamellar sheets. We approach it by building another geometric model from tubes and sheets.  Such an arrangement is also triply periodic; therefore, we fix a unit cell, which is not necessarily a minimal unit cell, with geometry that spans the whole structure. In the fixed unit cell, we puncture four holes per tube: one hole between a sheet that touches the tube from the above, and one from below, also one hole between two touching tubes on one side, and on the other side, see Figure \ref{fig:real_model}. Our model of such a geometric arrangement, see Figure \ref{fig:real_model}, demonstrates that with this approach, it is difficult to achieve a minimal surface that spans three periodic directions, which leaves the question about the assembly of lamellar structures and tubules open. Approaches related to this have been used to construct disordered minimal surfaces with potential applications to sponge phases \cite{himmelmann2025amorphous}.

\begin{figure}[H]
  \centering
  \setlength{\abovecaptionskip}{4pt}
  \setlength{\belowcaptionskip}{4pt}
  \vspace{-4pt}
  \begin{subfigure}[b]{0.23\linewidth}
    \centering
    \includegraphics[width=\linewidth]{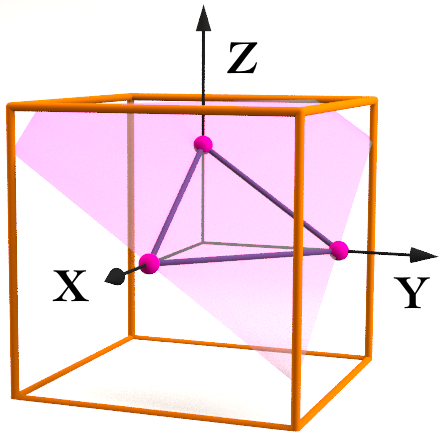}
    \caption{}
    \label{fig:hkl_111}
  \end{subfigure}\hspace{0.02\linewidth}%
  \begin{subfigure}[b]{0.23\linewidth}
    \centering
    \includegraphics[width=\linewidth]{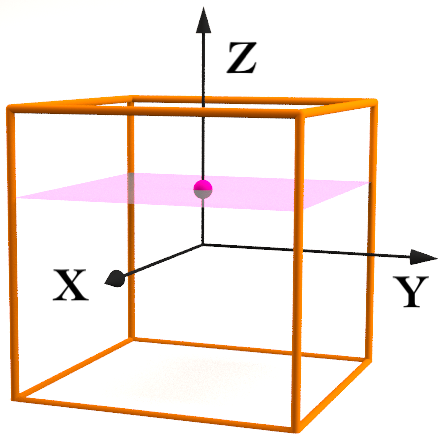}
    \caption{}
    \label{fig:hkl_001}
  \end{subfigure}\hspace{0.02\linewidth}%
  \begin{subfigure}[b]{0.23\linewidth}
    \centering
    \includegraphics[width=\linewidth]{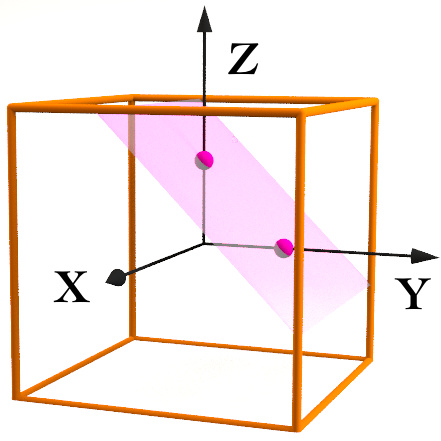}
    \caption{}
    \label{fig:hkl_011}
  \end{subfigure}
  \vspace{-4pt}
  \caption{ (a) Definition of Miller indices that define the orientation of slices. The triplet (hkl) denotes an orientation that corresponds to a plane that intersects the (xyz)-axes at the points $\nicefrac{1}{h}$, $\nicefrac{1}{k}$ and $\nicefrac{1}{l}$. Since here we have the case of a cubic lattice, the plane defined by (hkl) is orthogonal to $[h, k, l]$ vector. In this particular case, the orientation plane corresponds to the triple (111). (b) Orientation defined by (001). (c) Orientation defined by (011)}

  \label{fig:hkl}
\end{figure}

The formation of this suite of specific triply-periodic minimal surfaces from the cubic rod packings suggests the potential for these surfaces to be identified in cubic membranes at various stages of formation. As an initial exploration in this direction, we construct simulated TEM micrographs of these surfaces. Our procedure follows the approach implemented in SPIRE \cite{hain2022spire} for generating 2‑D slices. Extracting slices from each surface at various orientations yields a gallery of possible 2‑D patterns (Table \ref{Tab:TEM}). Here, an orientation is defined by a set of Miller indices $(h,k,l)$ \cite{kittel2018introduction}, as shown in Figure \ref{fig:hkl}. In general, Miller indices define families of parallel lattice planes; here, we consider a cubic lattice defined by orthogonal unit vectors. A plane from such a family intersects the $(x,y,z)$ axes at the points $\nicefrac{1}{h}$, $\nicefrac{1}{k}$, and $\nicefrac{1}{l}$, illustrated by pink spheres in Figure \ref{fig:hkl}. Each simulated surface is first oriented with respect to the chosen Miller indices, then a slice of size $400 \times 400 \times 70$ nm is cut out containing 4 unit cells. To imitate a TEM image, we sum the voxels along the thinnest dimension and normalise the resulting array to obtain a grayscale image. We then searched for $(hkl)$ triples that produce the least regular patterns, as such patterns are most often observed in membrane TEM images.

The patterns summarised in Table \ref{Tab:TEM} demonstrate possible slice patterns for identification of sponge or intermediate phases (Figure \ref{fig:tubules_figure}a in developing PLBs. Among the surfaces considered, $C(^\pm Y)$ and $C(I_2-Y^{**})$ generate the most realistic projections, providing slices that are qualitatively similar to the experimental data presented on the right side of Figure \ref{fig:tubules_figure}a. These projections expand the set of plausible geometric templates for biological cubic membranes and support the idea that multiple TPMS type, beyond the Diamond, may transiently arise during membrane reorganisation.

\begin{table}[htbp]
  \centering
  \renewcommand{\arraystretch}{1.5}
  \setlength{\tabcolsep}{4pt}
  \begin{tabular}{c  c  c  c  c  c }
     \(\mathit{hkl}\) & \(I\!-\!WP\) & \(C(\pm Y)\) & \(H\)
                   & \(C(I_{2}-Y^{**})\) & \makecell{Tubuls and sheets\\ arrangment } \\
  \hline

      &
      \includegraphics[height=2.3cm]{surfaces_table/iwp2.png} &
      \includegraphics[height=2.3cm]{surfaces_table/cy_darker.png} &
      \includegraphics[height=2.3cm]{surfaces_table/omega_h2.png} &
      \includegraphics[height=2.3cm]{surfaces_table/gamma.png}  &
      \includegraphics[height=2.3cm]{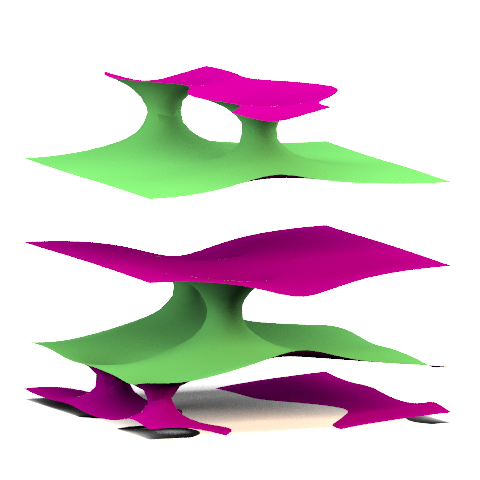} \\[3pt]
    
    1 0 0 &
      \includegraphics[height=2.3cm]{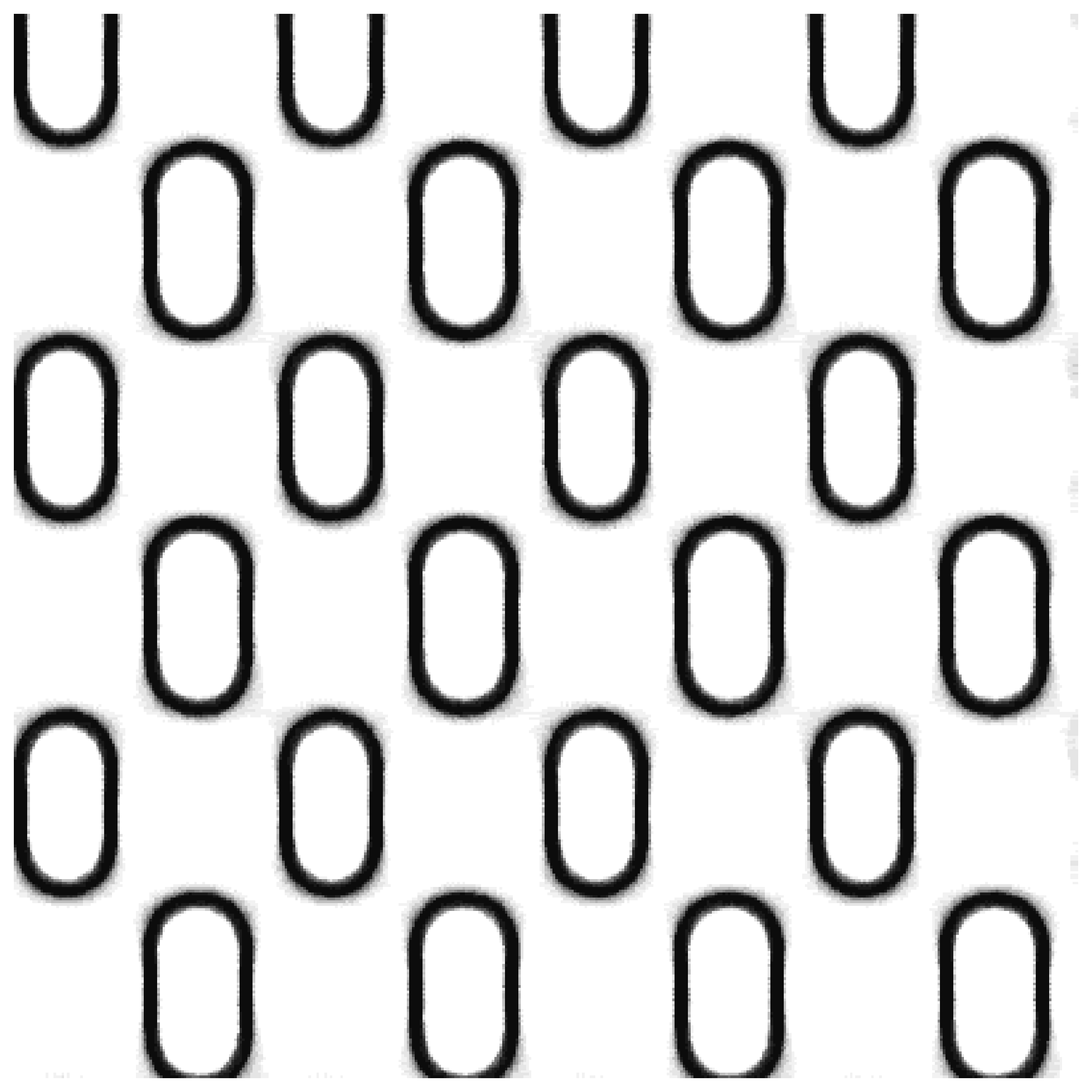} &
      \includegraphics[height=2.3cm]{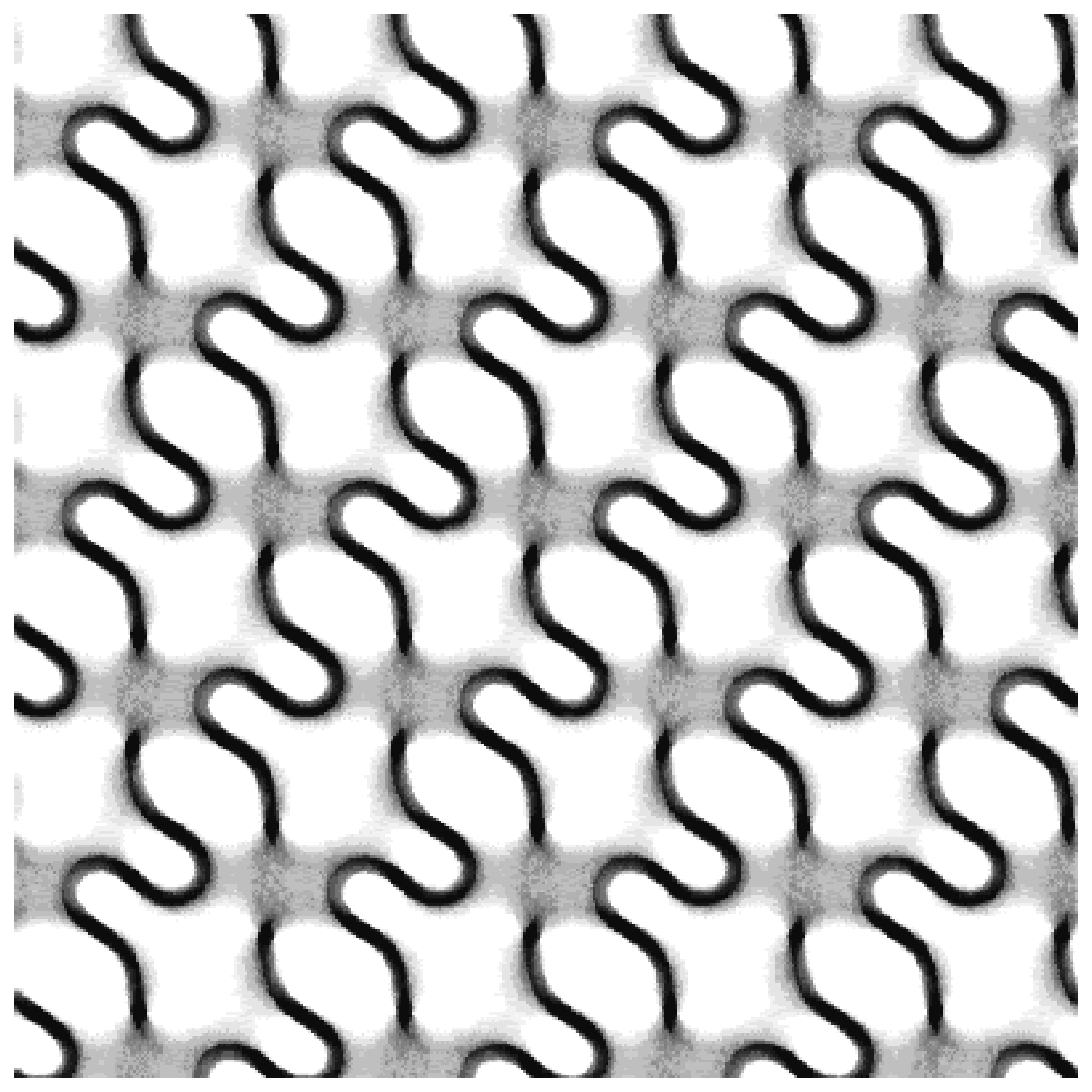} &
      \includegraphics[height=2.3cm]{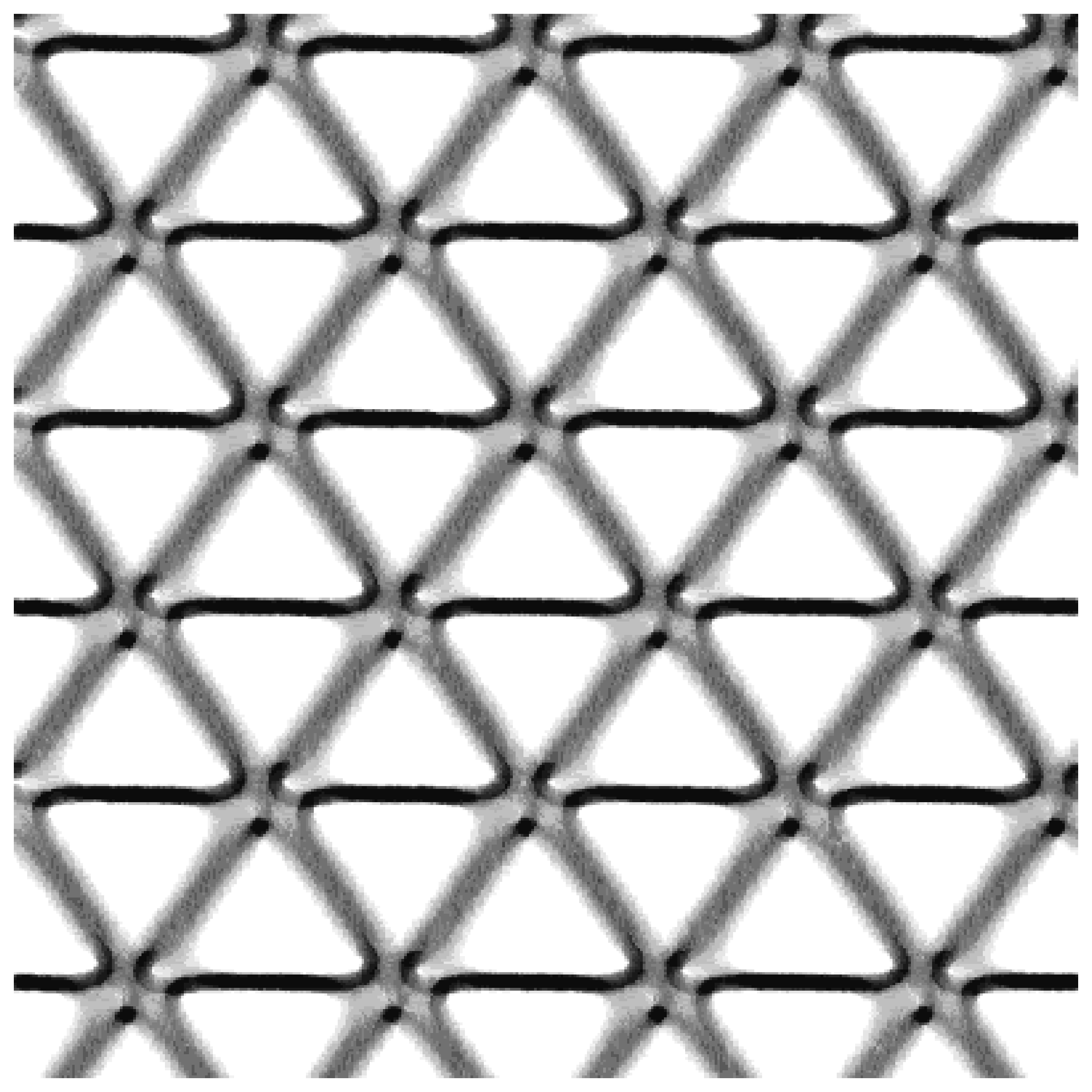} &
      \includegraphics[height=2.3cm]{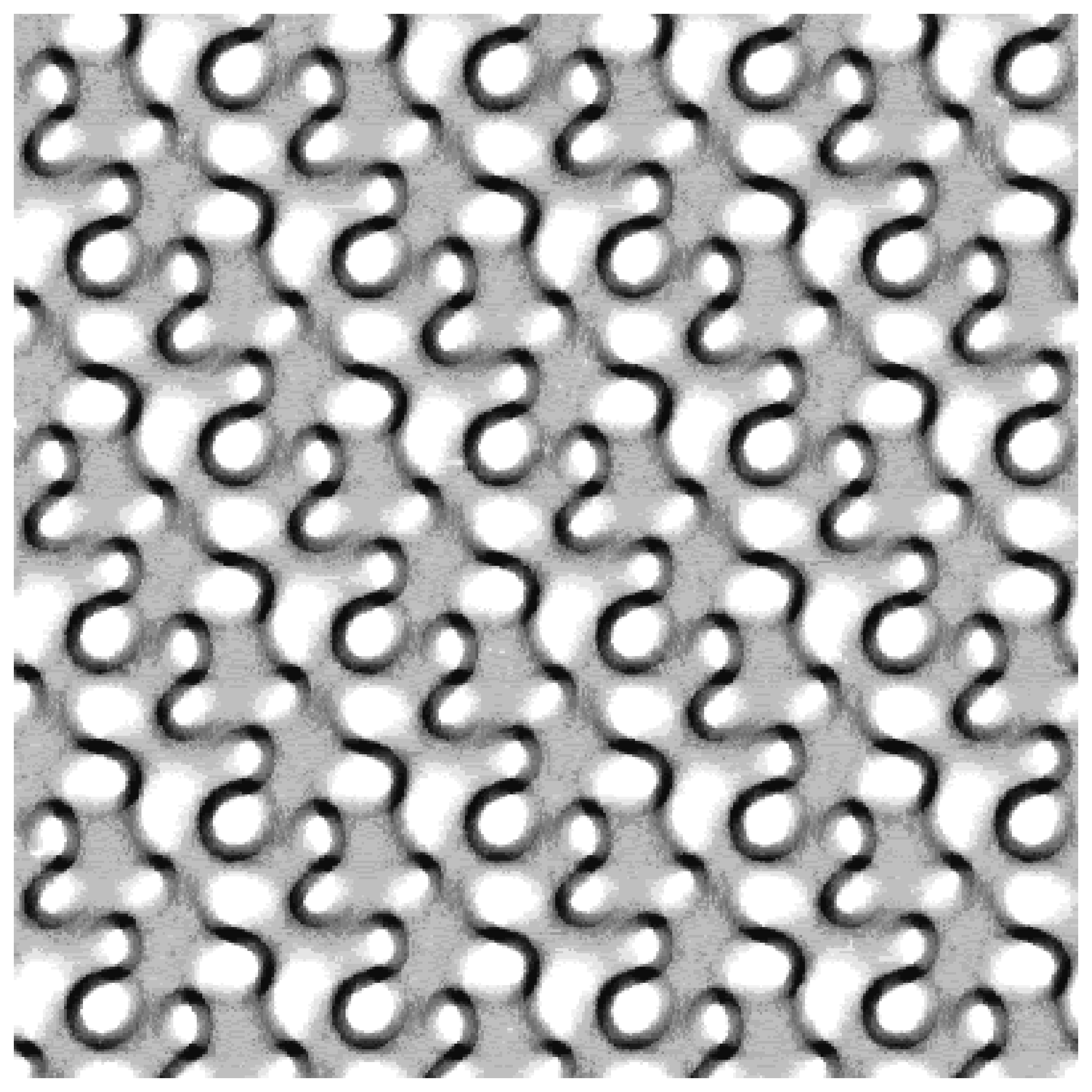}  &
      \includegraphics[height=2.3cm]{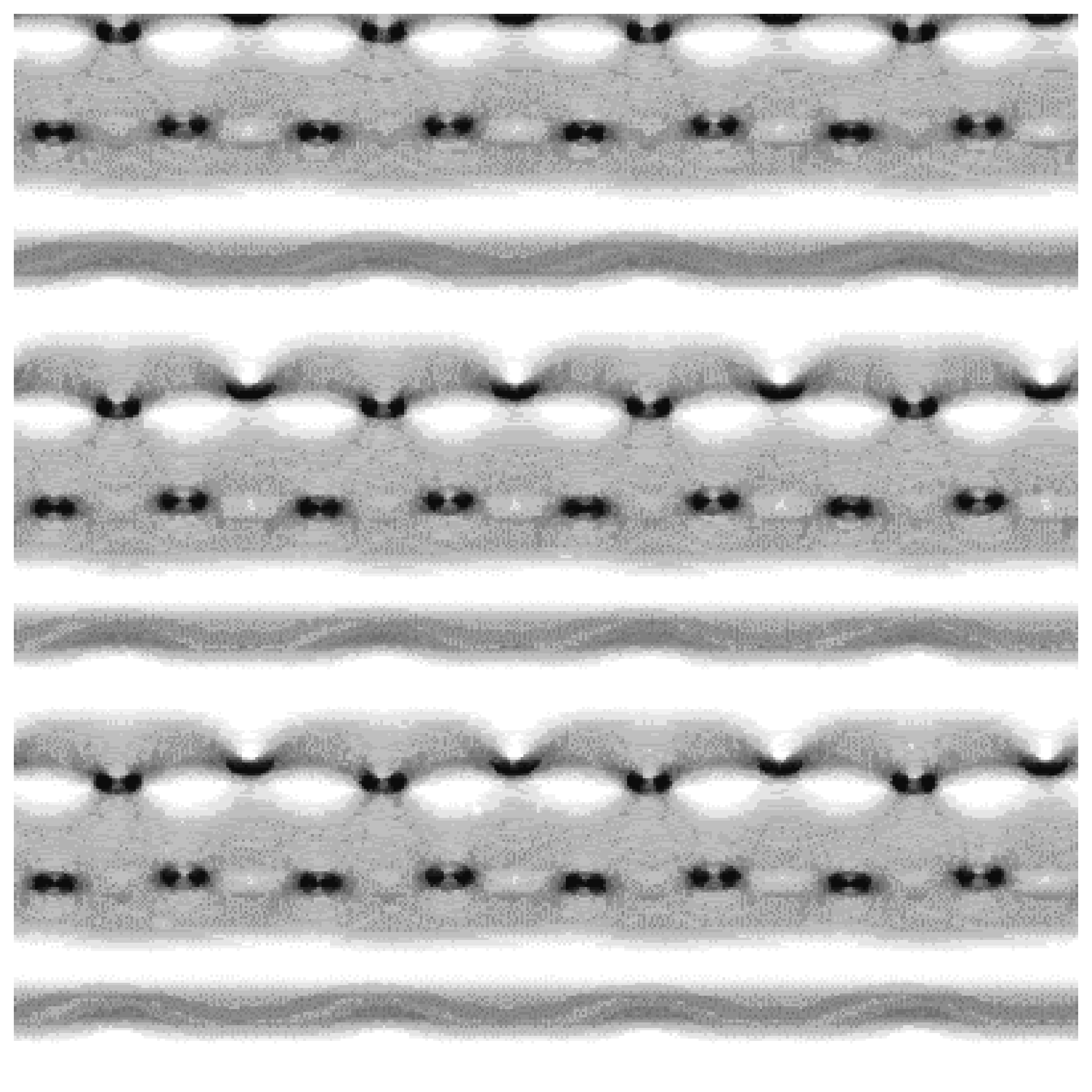} \\[3pt]

    1 0 1 &
      \includegraphics[height=2.3cm]{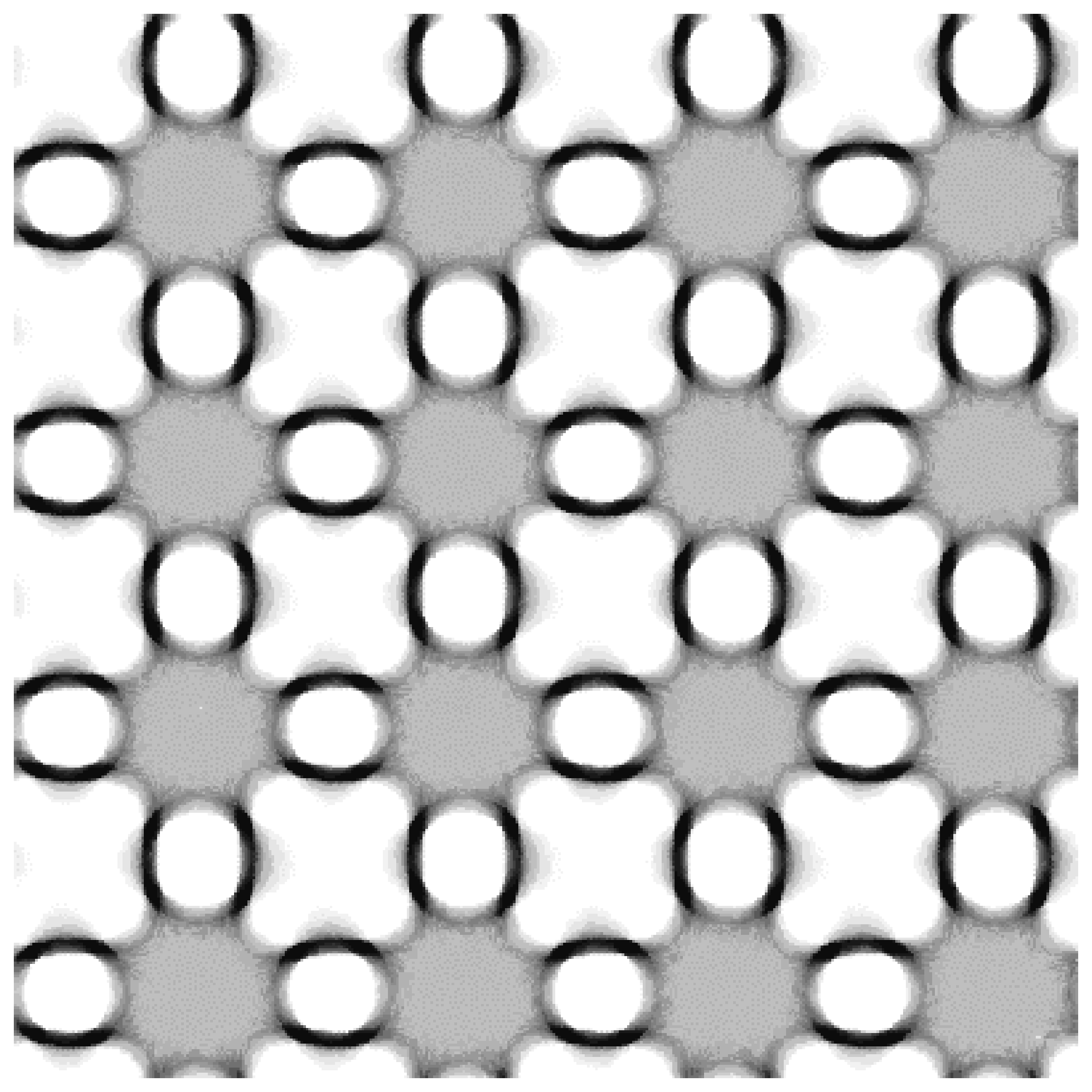} &
      \includegraphics[height=2.3cm]{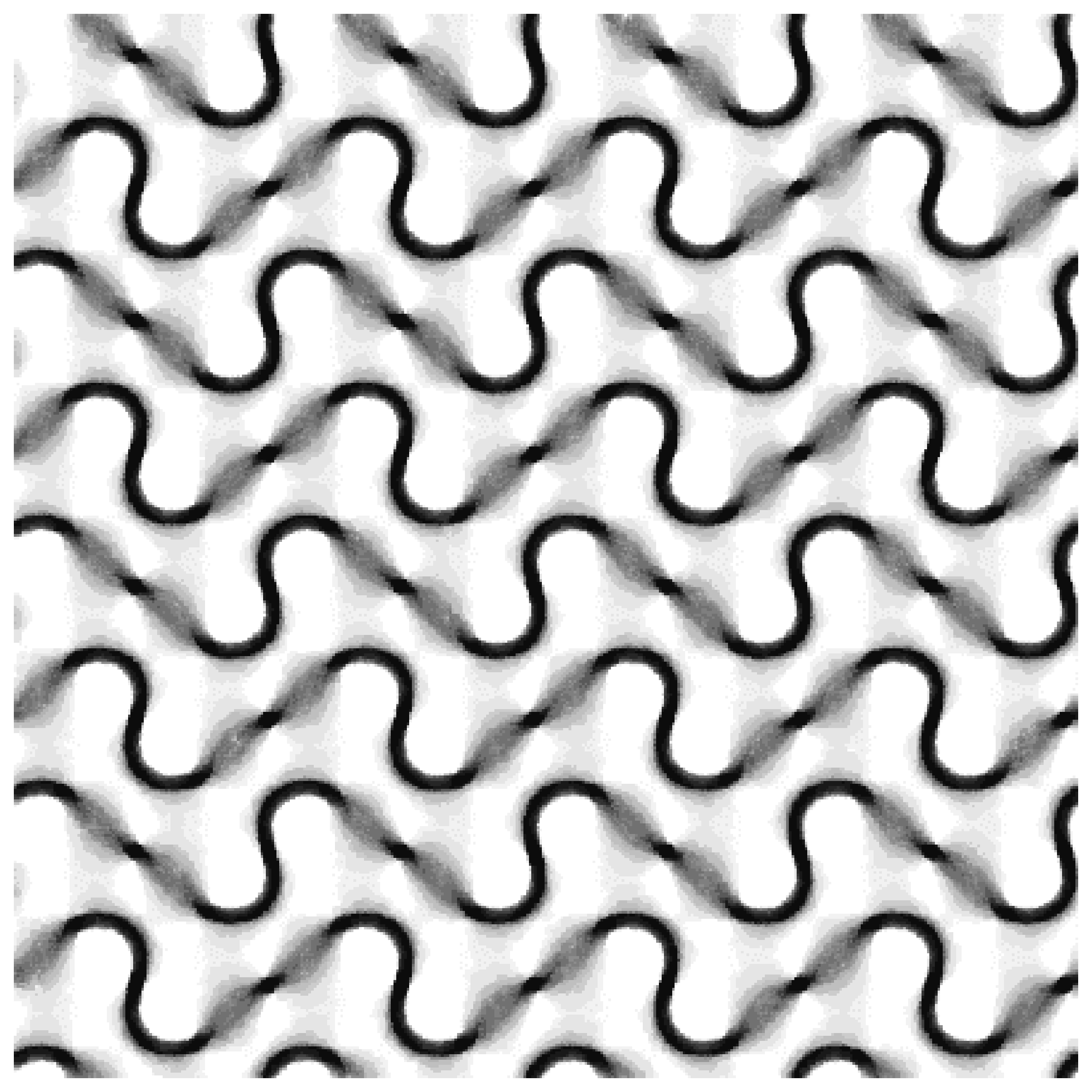} &
       \includegraphics[height=2.3cm]{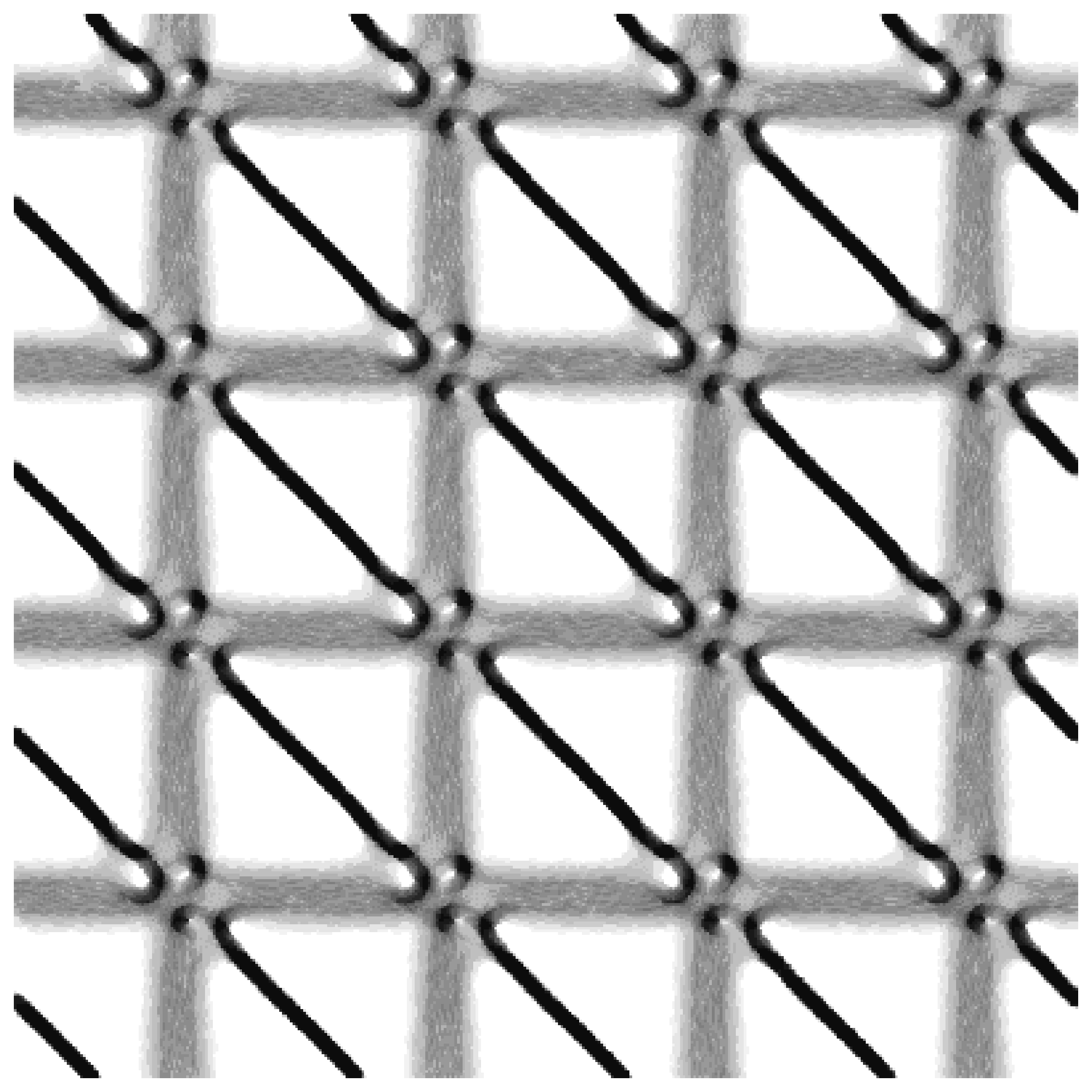} &
      \includegraphics[height=2.3cm]{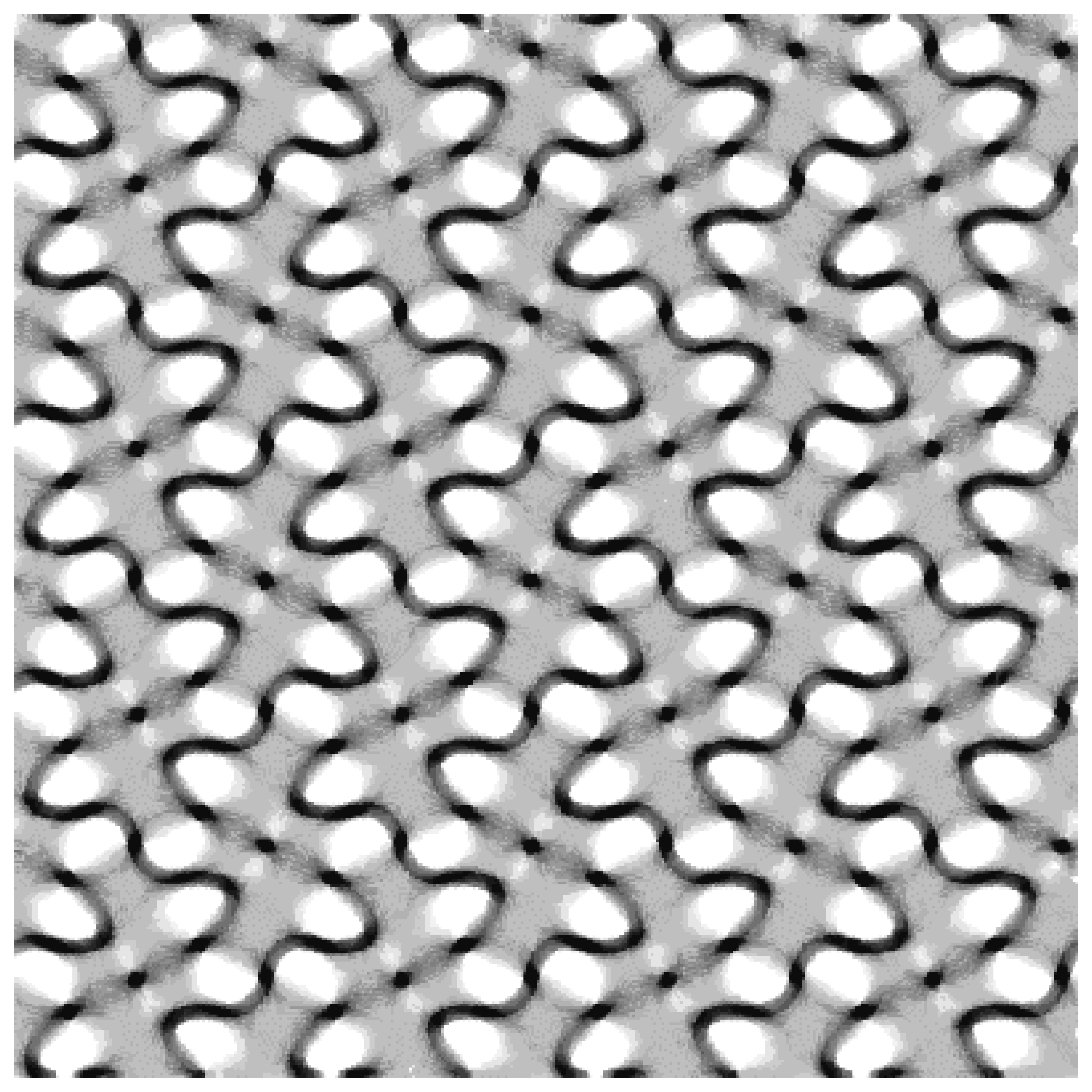} &
       \includegraphics[height=2.3cm]{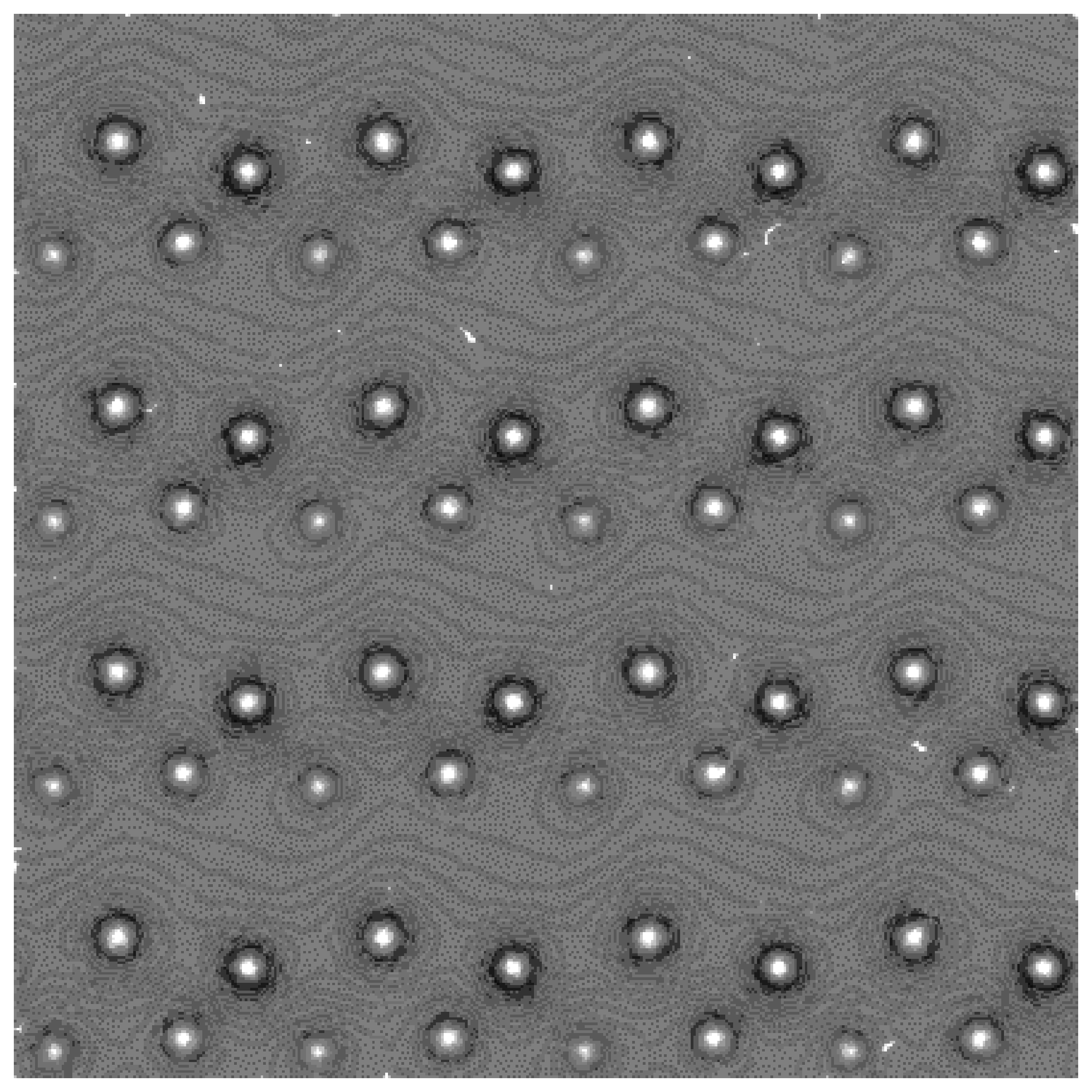} \\[3pt]

    1 1 0 &
      \includegraphics[height=2.3cm]{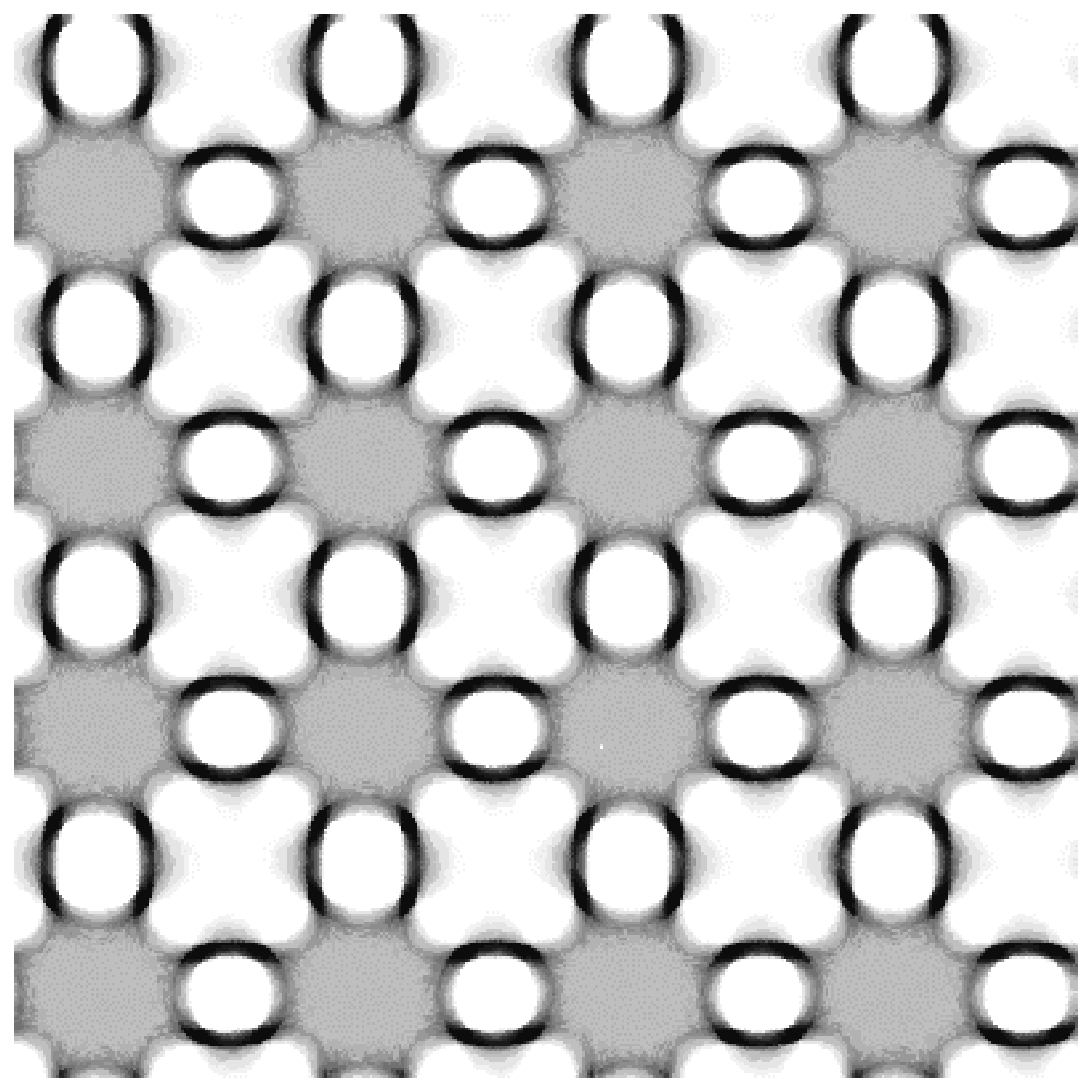} &
      \includegraphics[height=2.3cm]{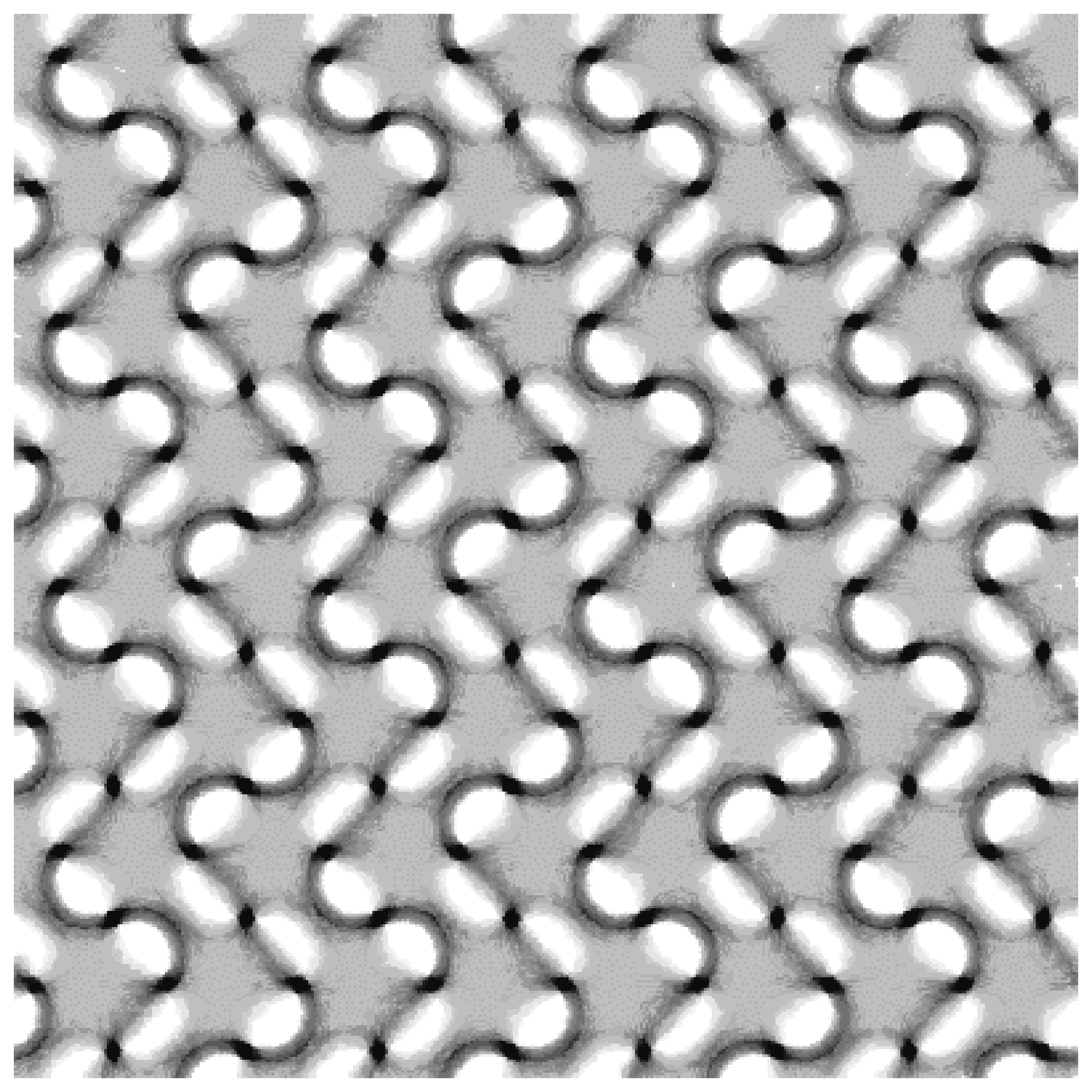}&
     \includegraphics[height=2.3cm]{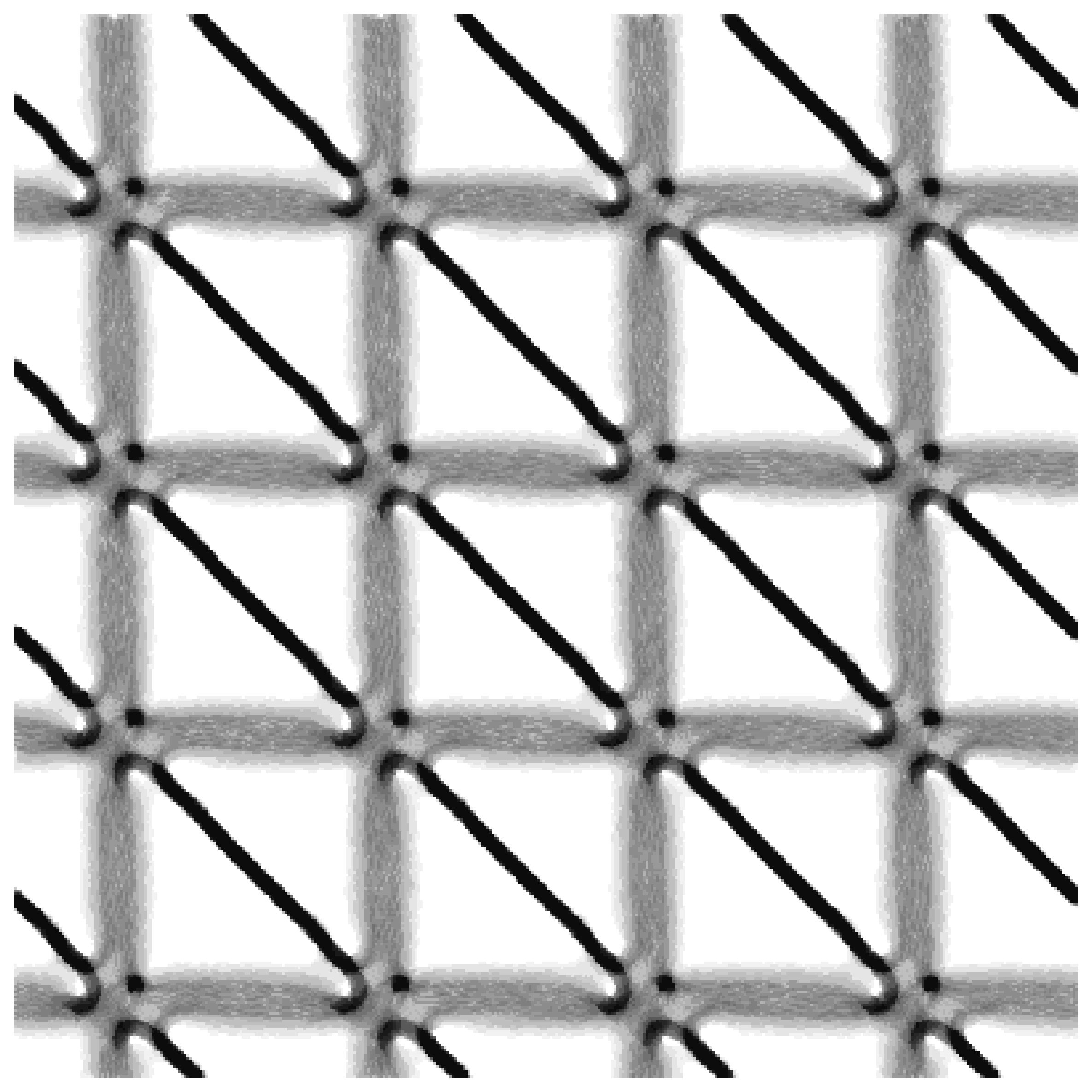} &
       \includegraphics[height=2.3cm]{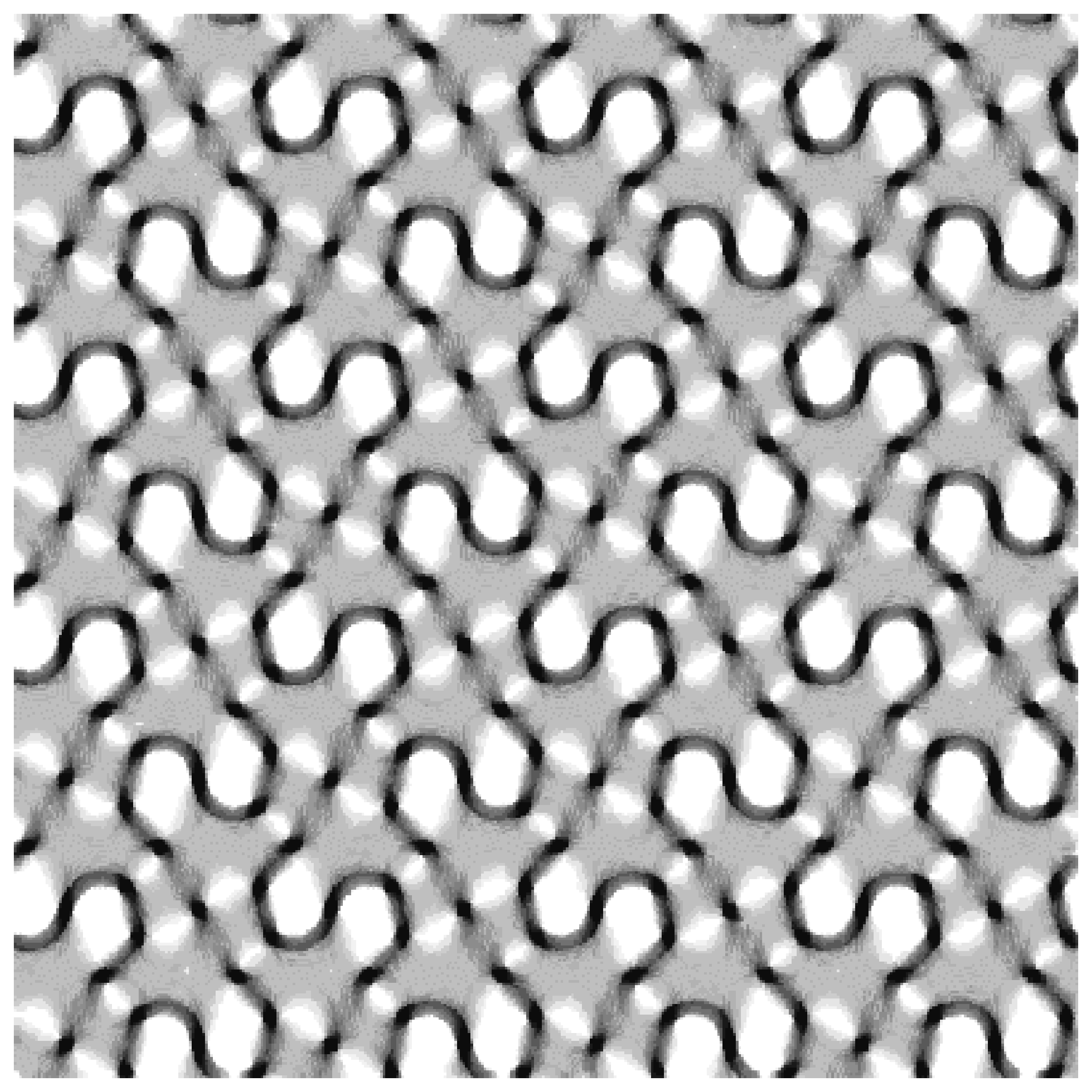} &
      \includegraphics[height=2.3cm]{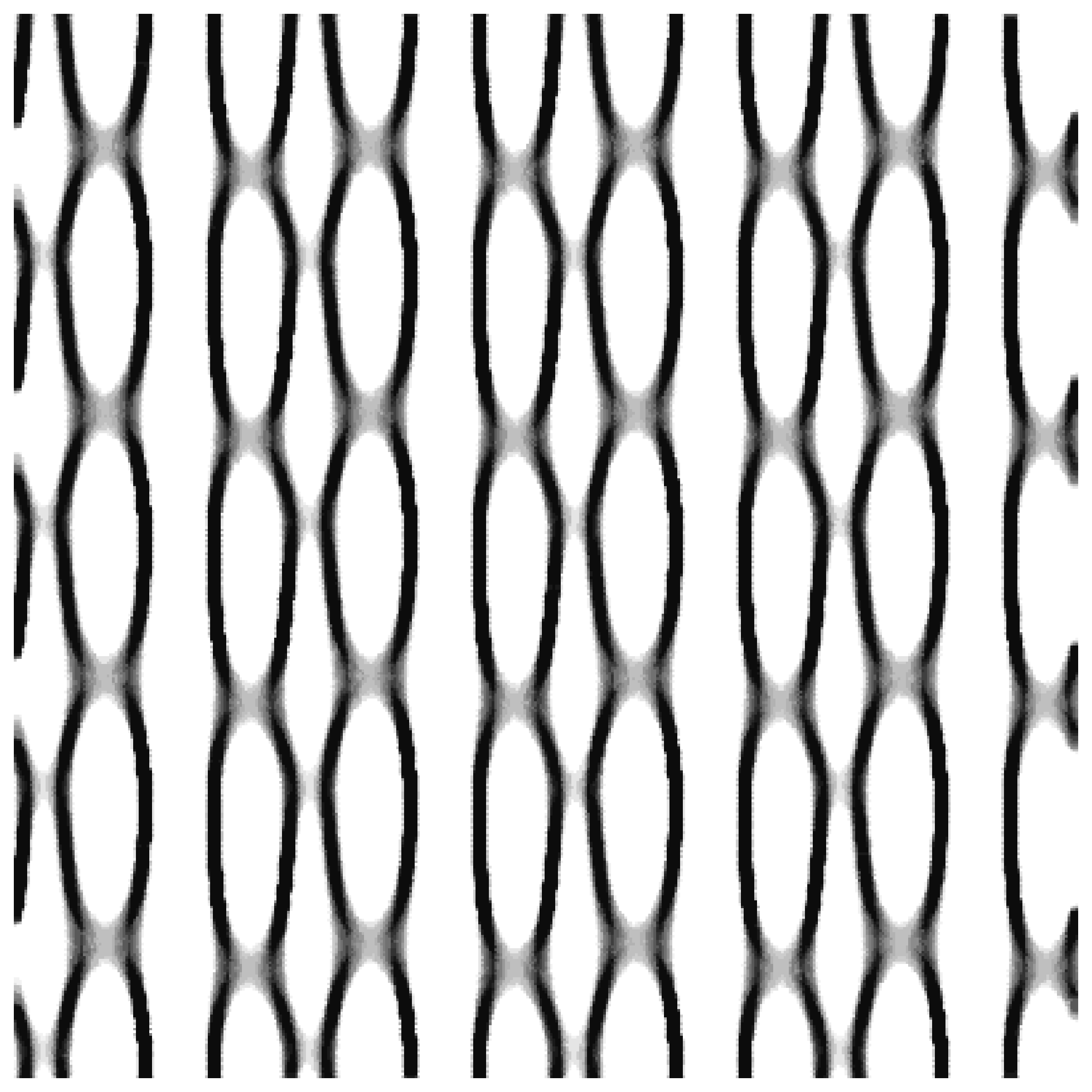}\\[3pt]

    1 1 1&
      \includegraphics[height=2.3cm]{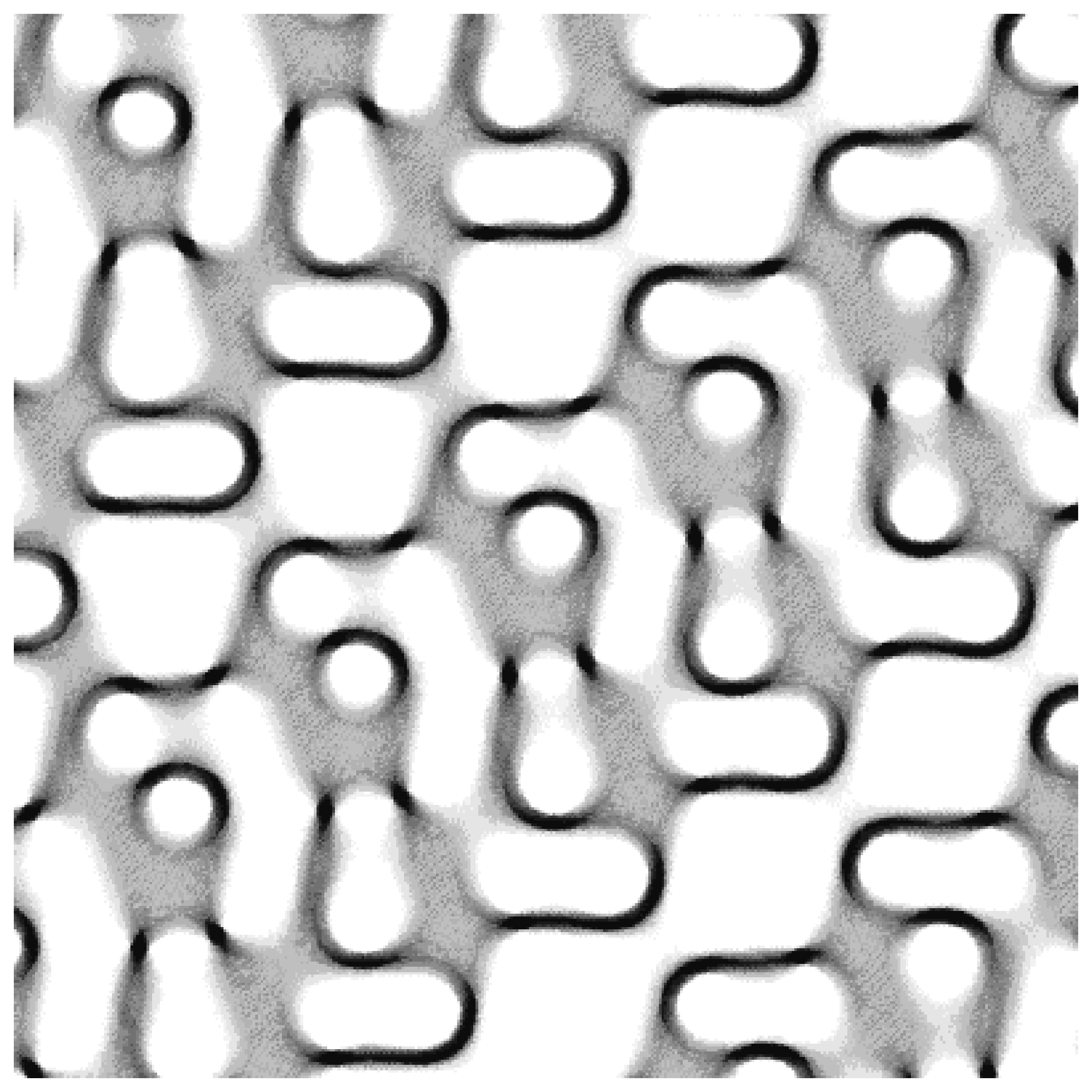} &
      \includegraphics[height=2.3cm]{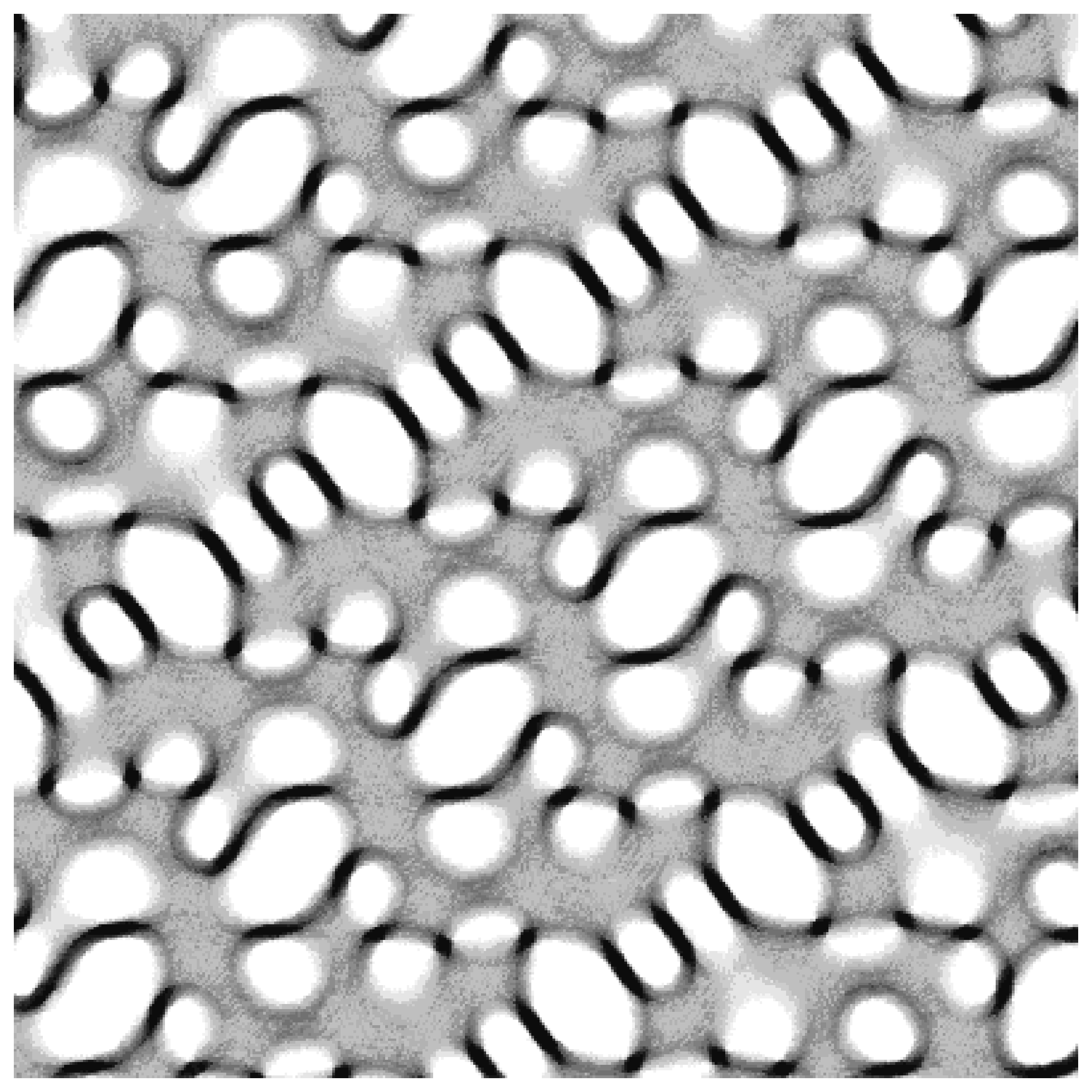} &
      \includegraphics[height=2.3cm]{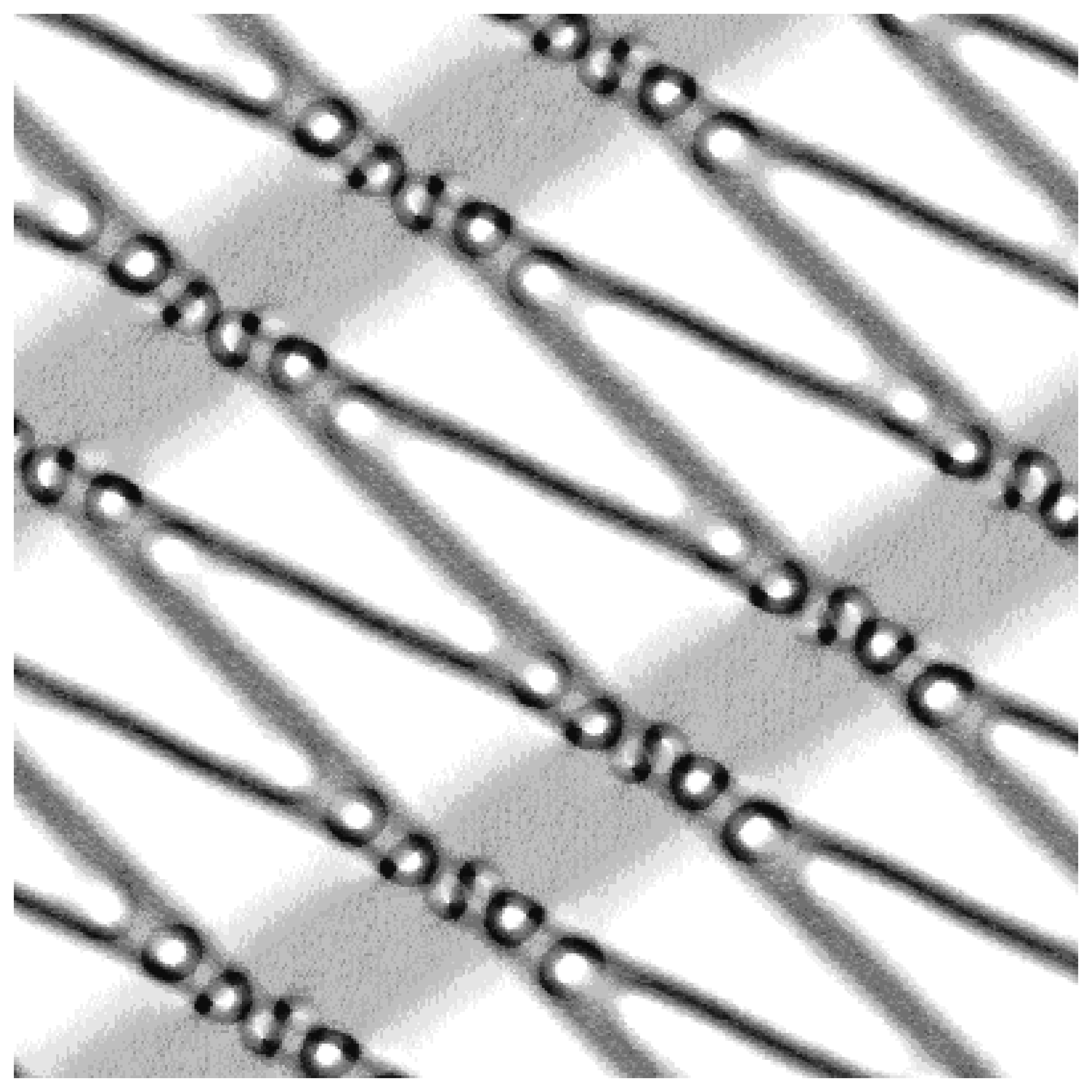} &
      \includegraphics[height=2.3cm]{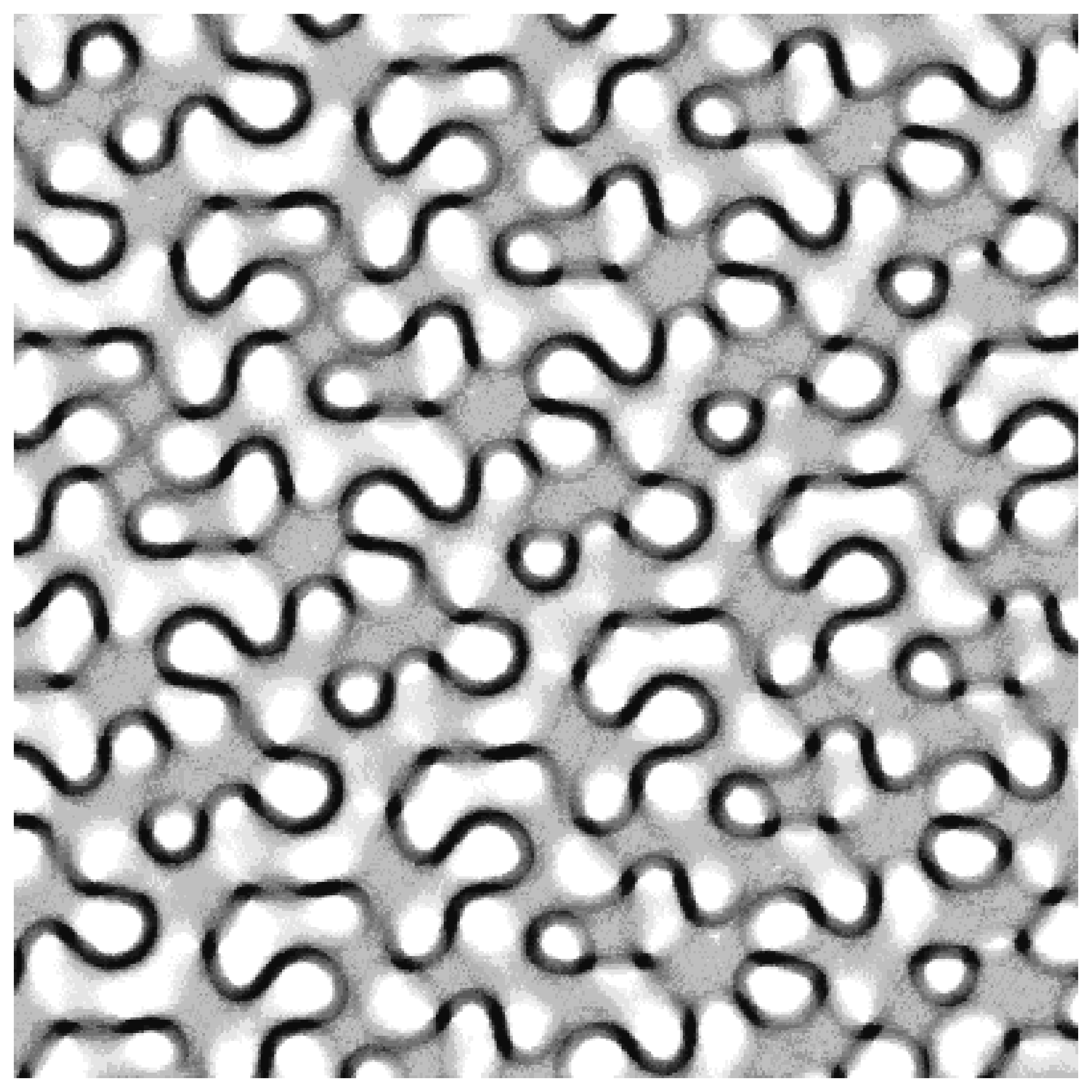} &
      \includegraphics[height=2.3cm]{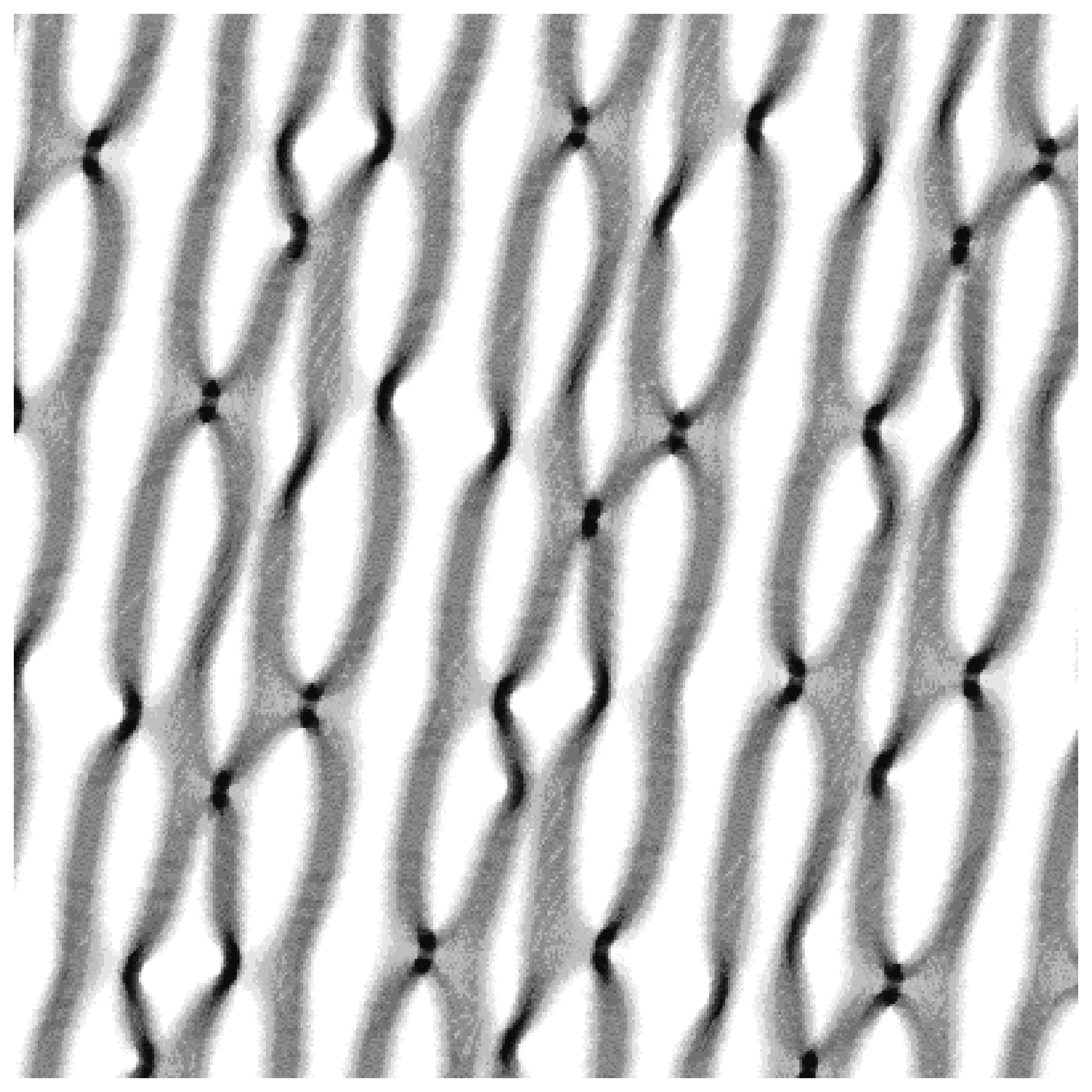} \\[3pt]

    9 4 7&
      \includegraphics[height=2.3cm]{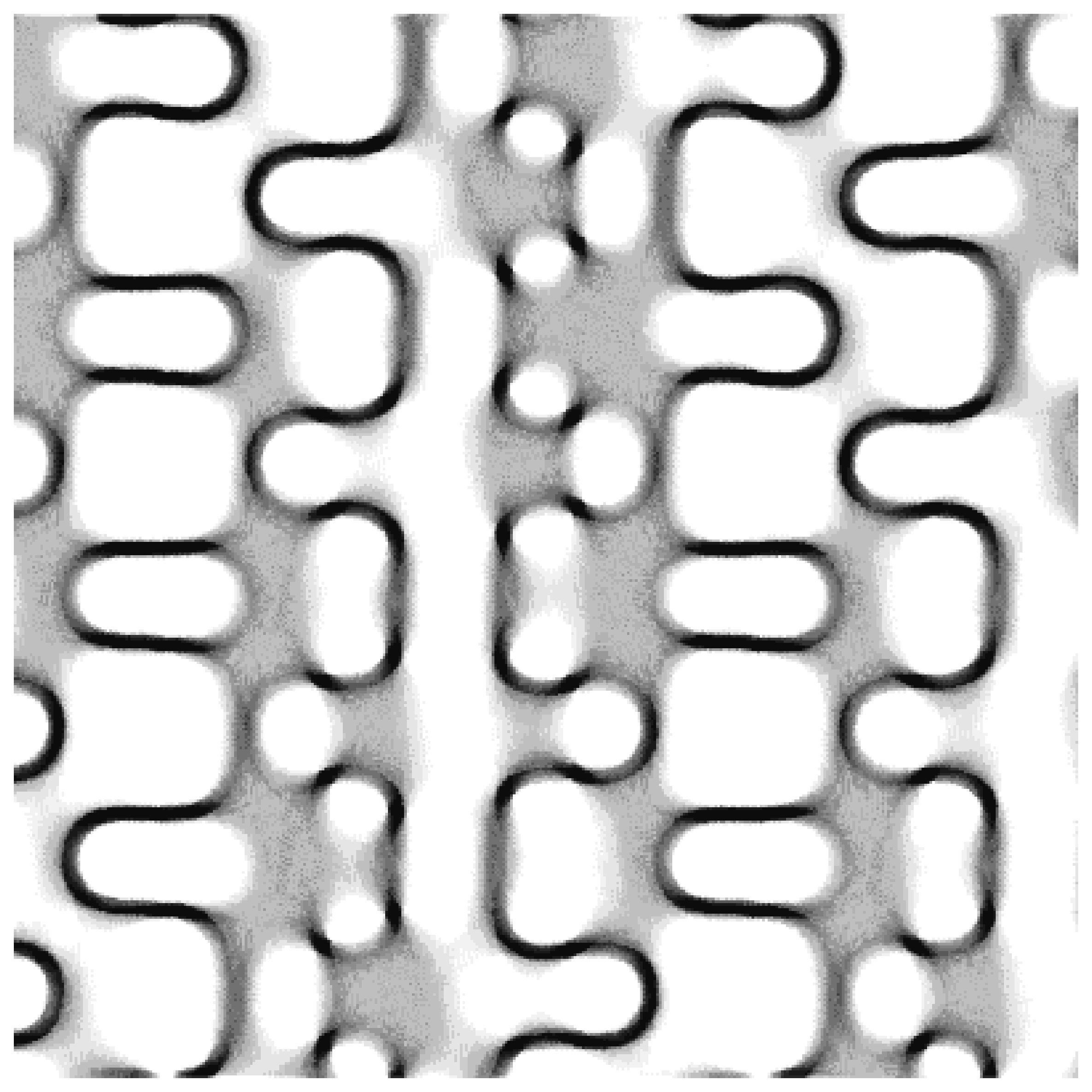} &
      \includegraphics[height=2.3cm]{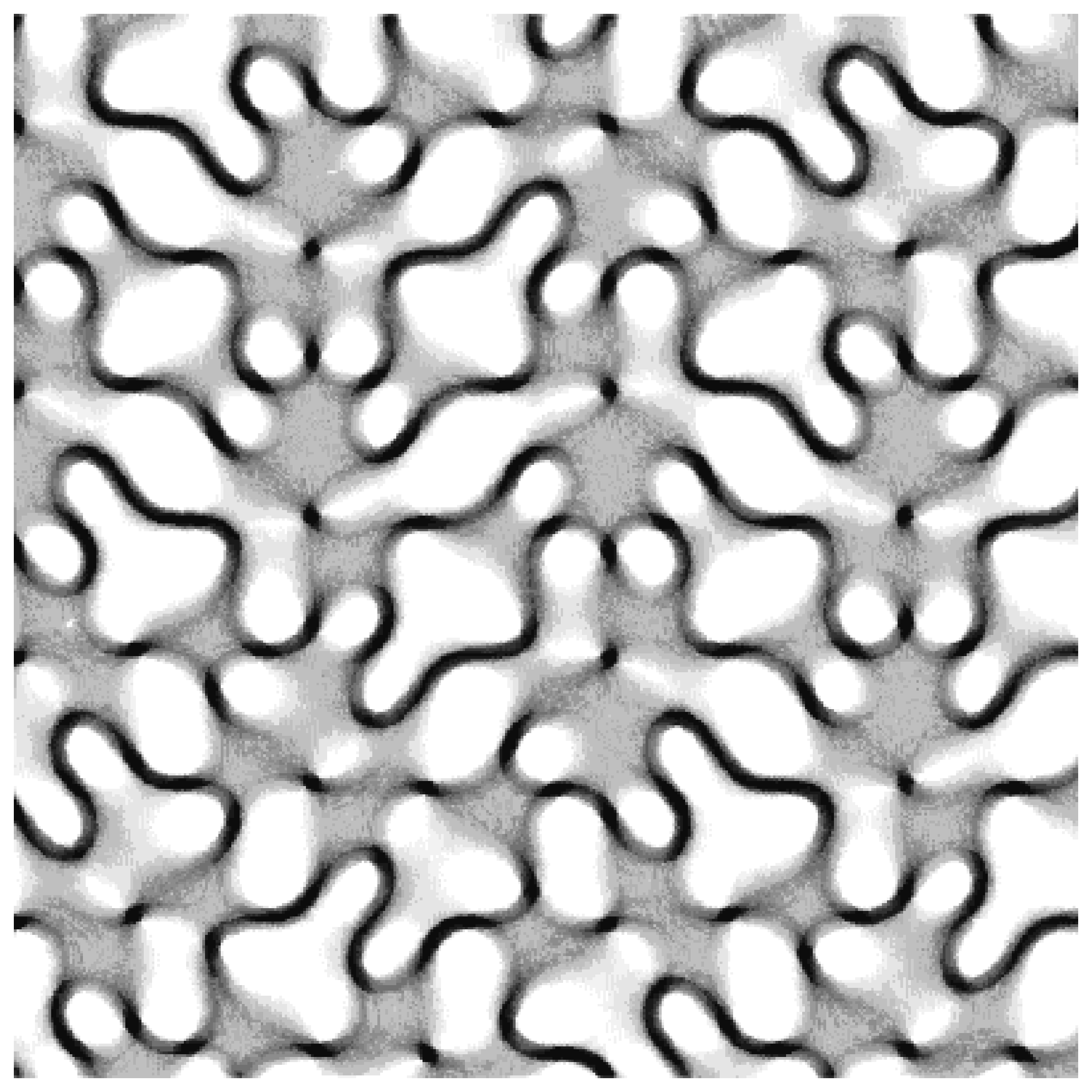}  &
      \includegraphics[height=2.3cm]{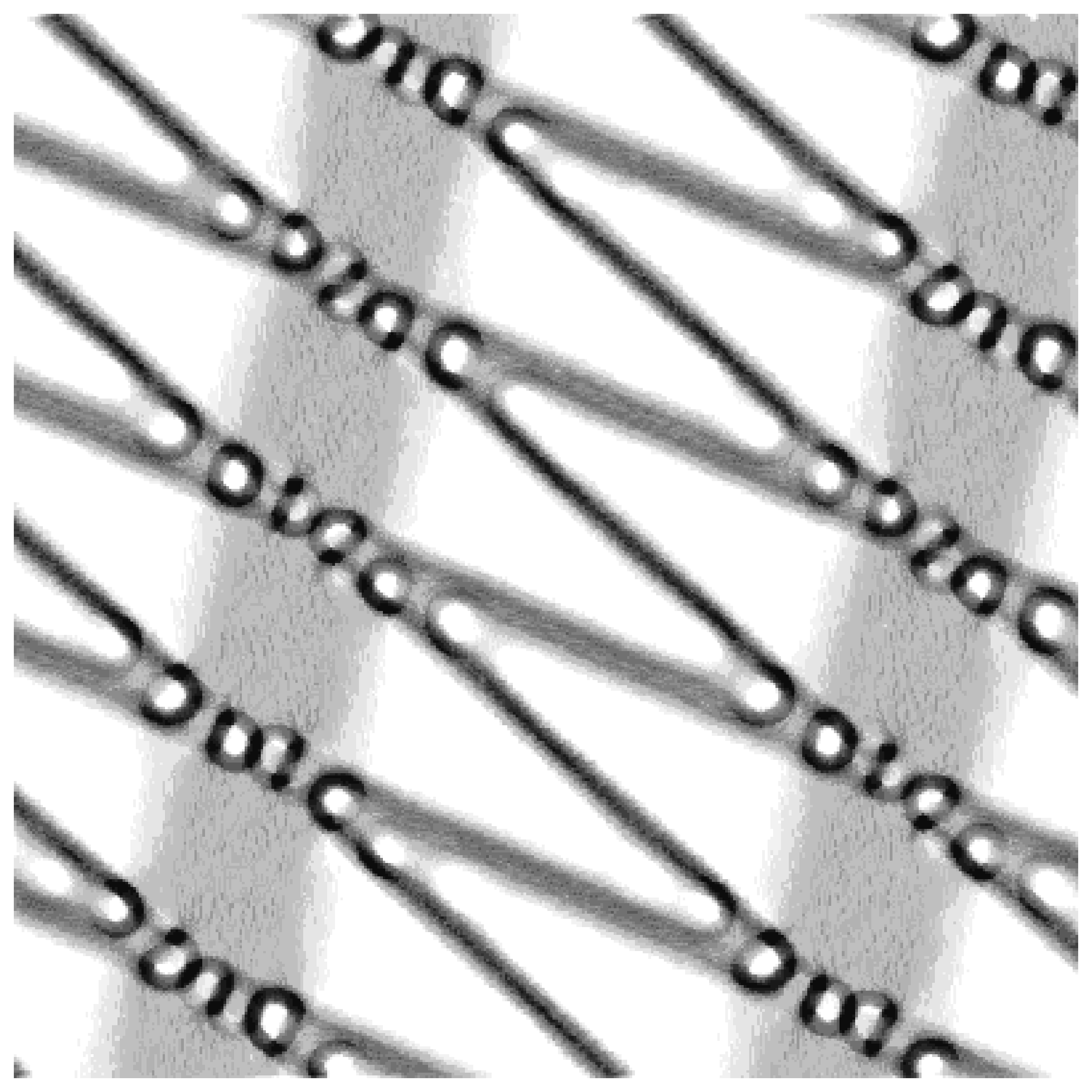} &
     \includegraphics[height=2.3cm]{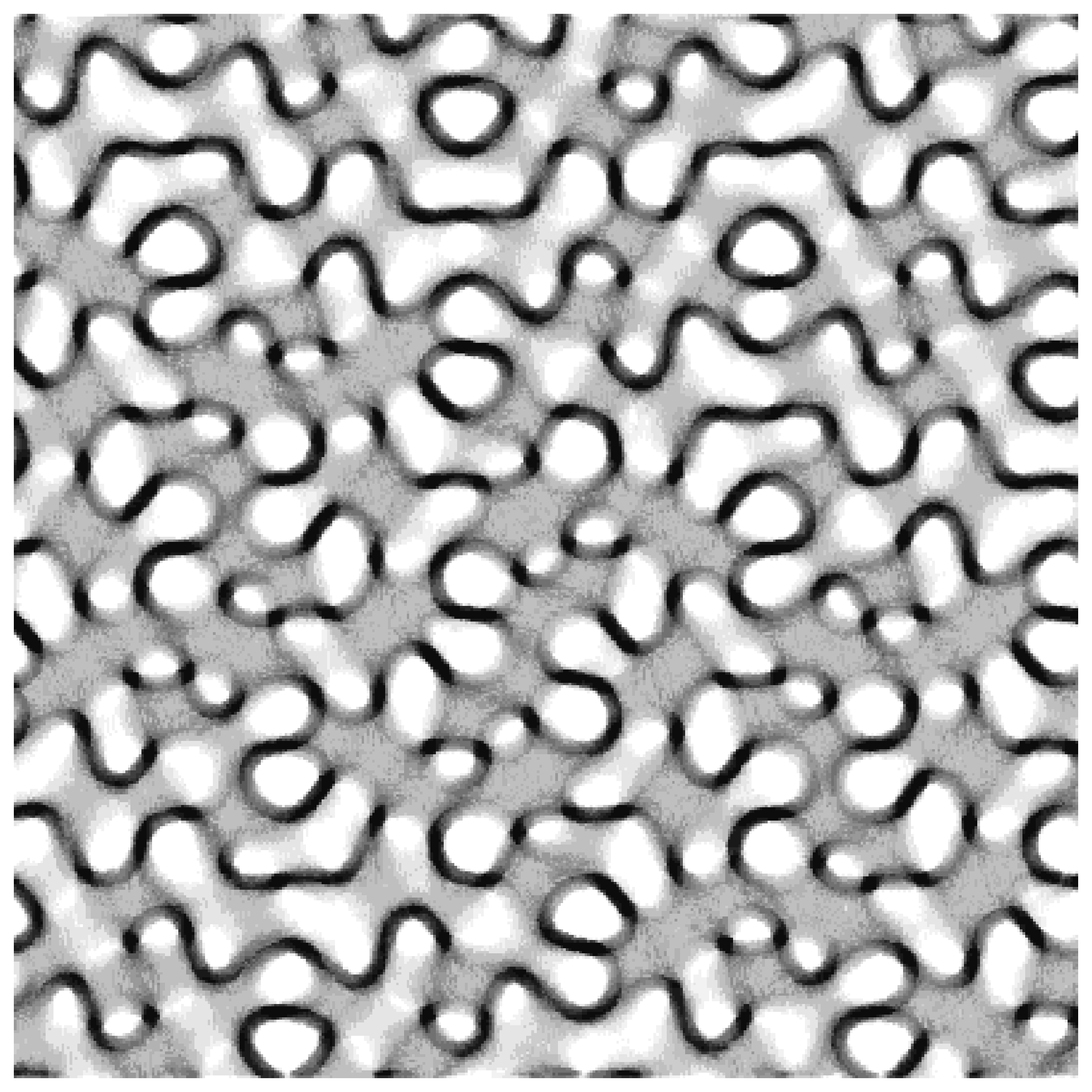} &
      \includegraphics[height=2.3cm]{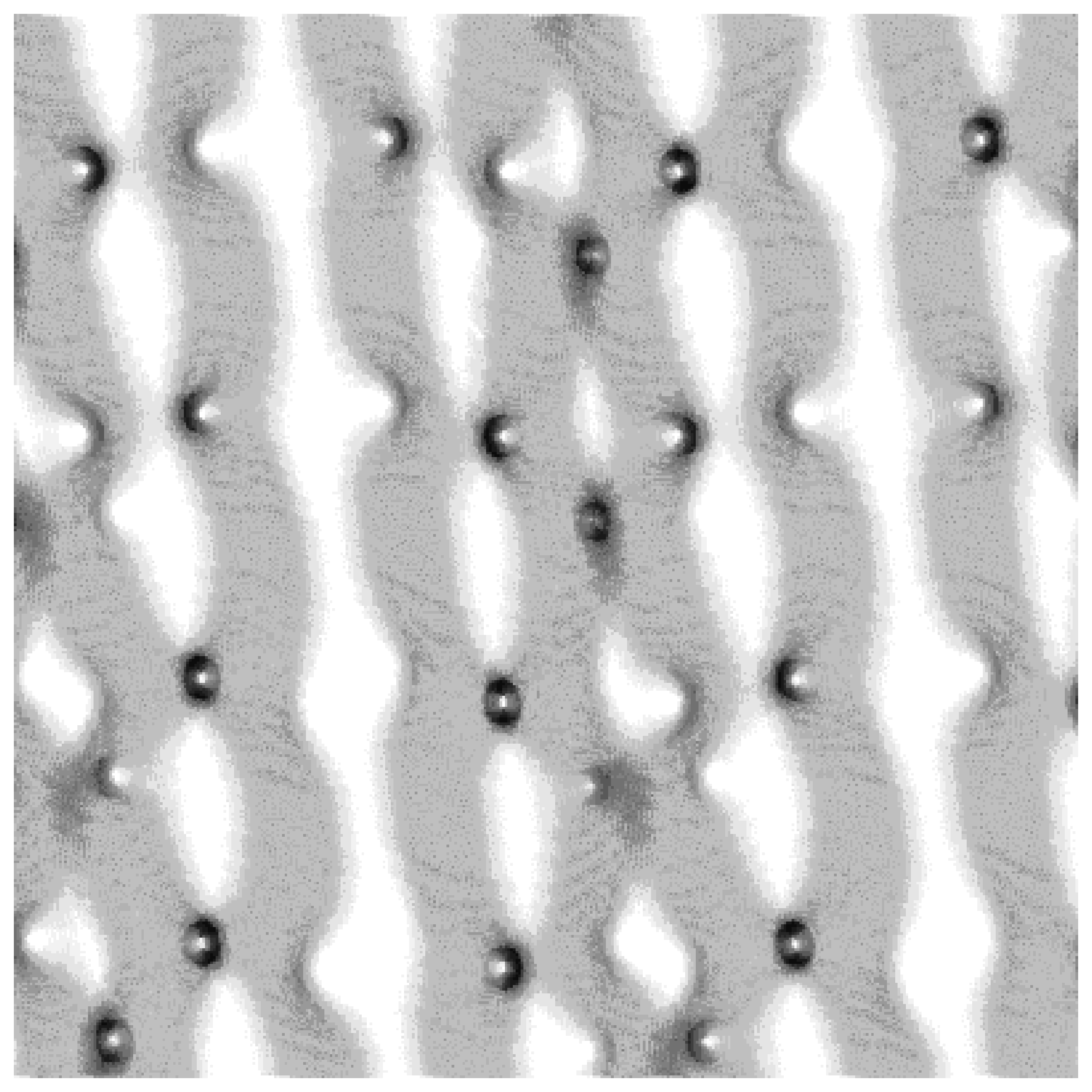} \\[3pt]

    3 2 1 &
      \includegraphics[height=2.3cm]{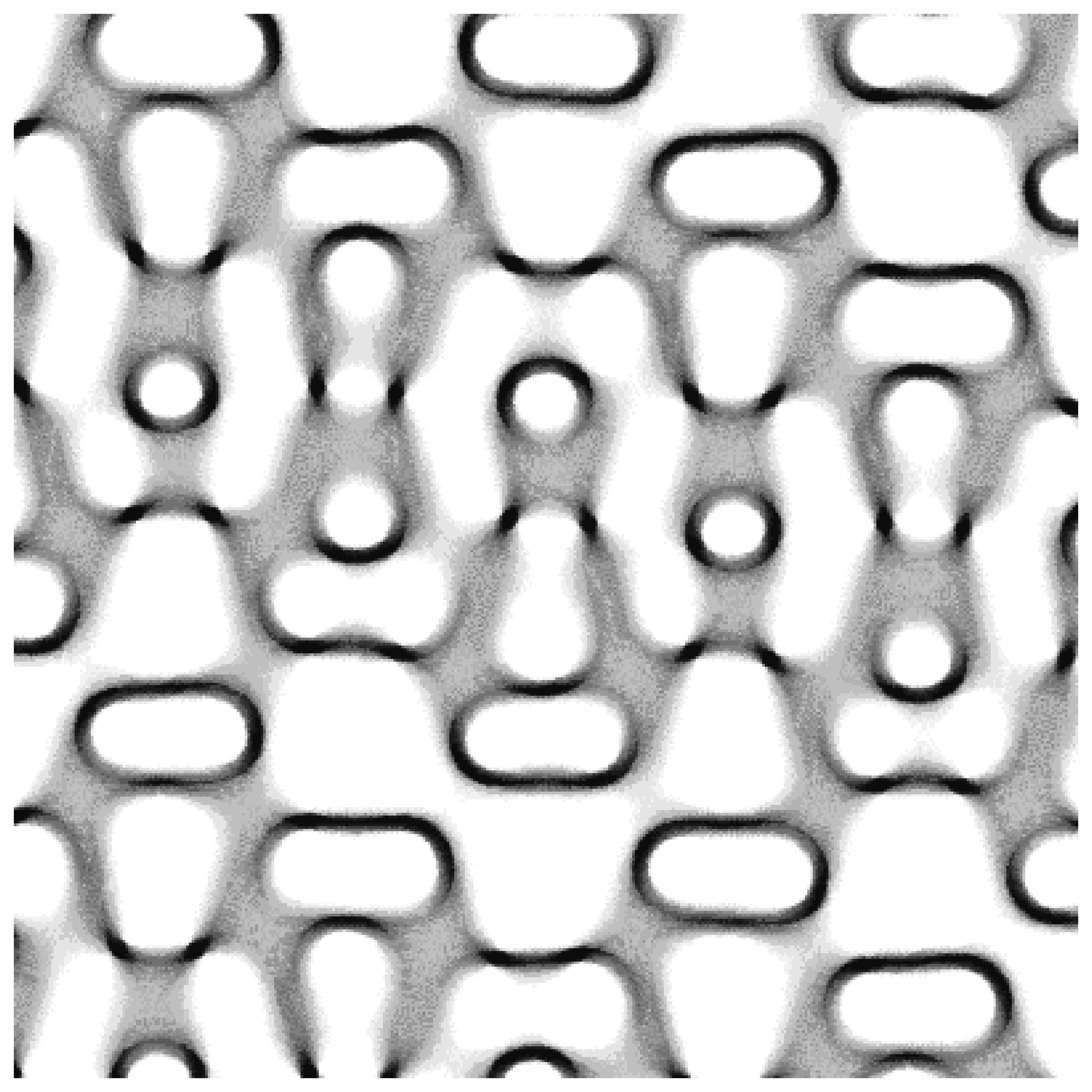} &
      \includegraphics[height=2.3cm]{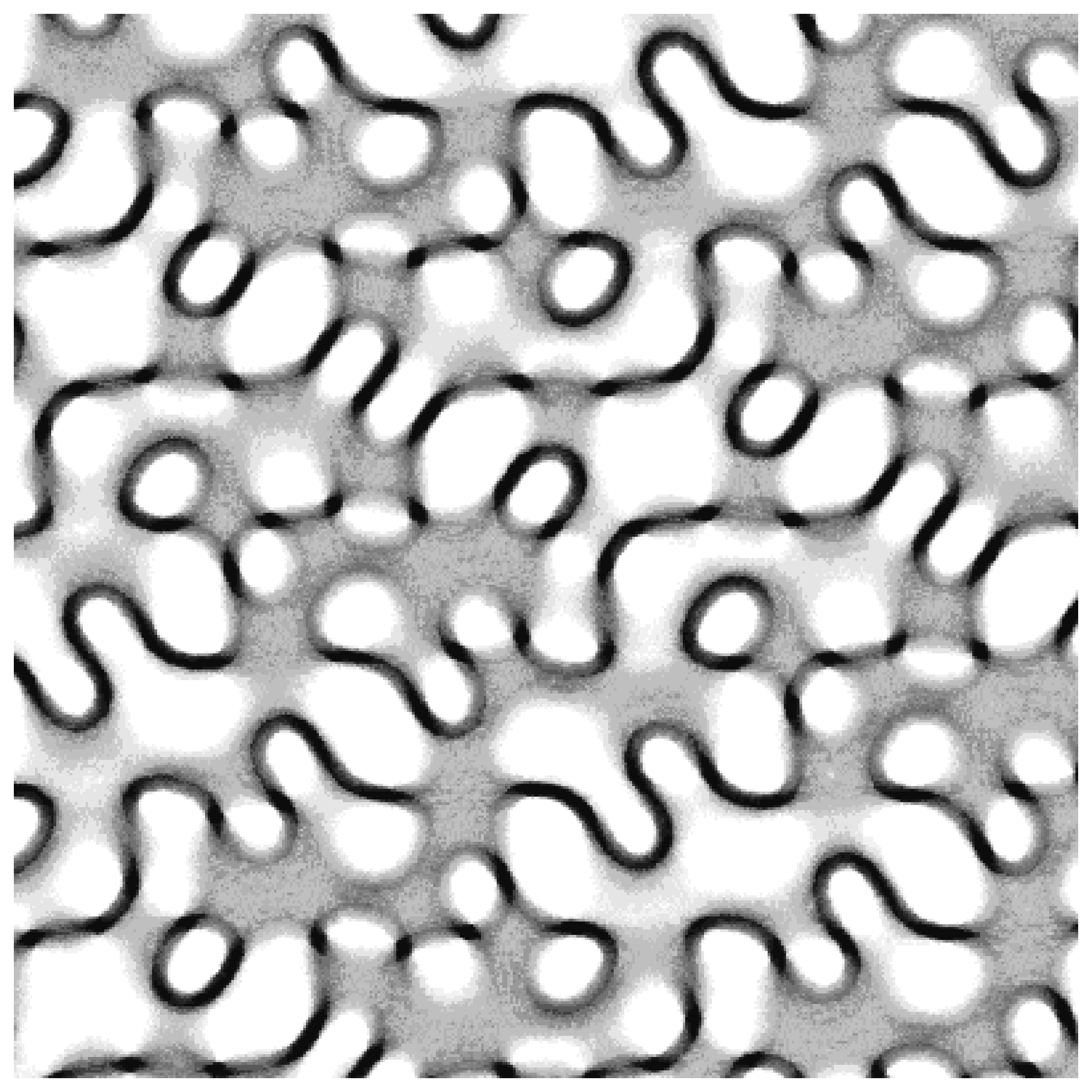} &
       \includegraphics[height=2.3cm]{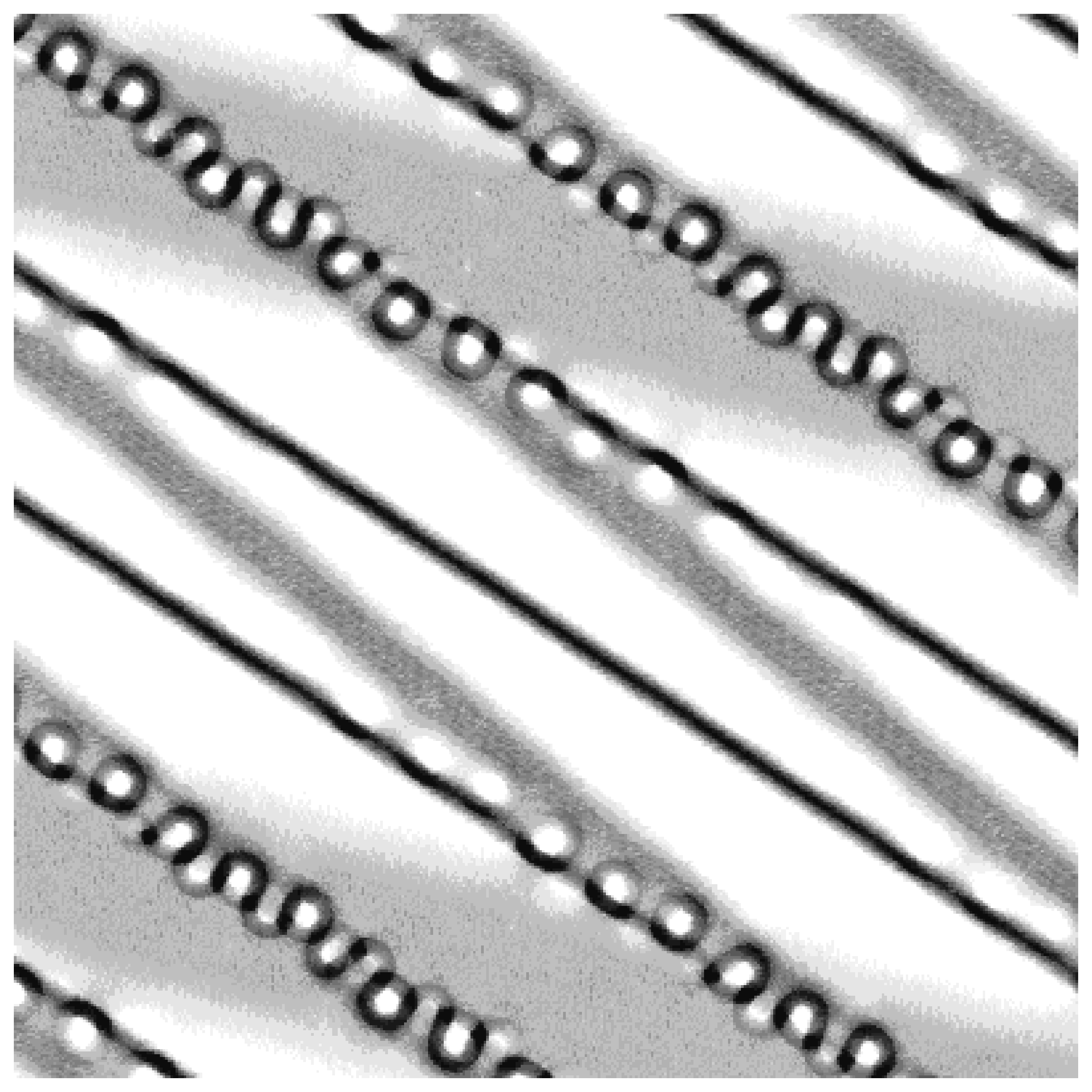} &
      \includegraphics[height=2.3cm]{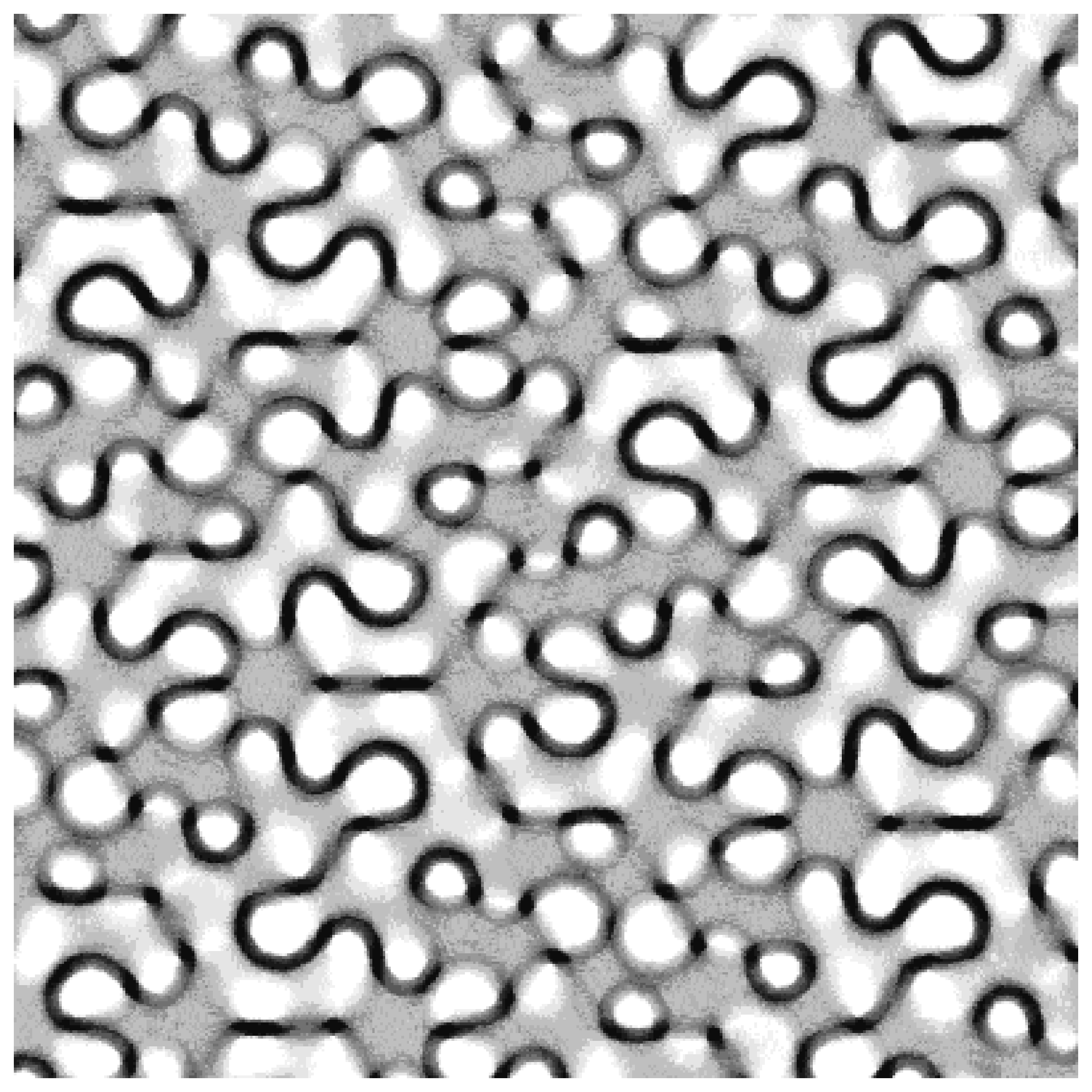} &
      \includegraphics[height=2.3cm]{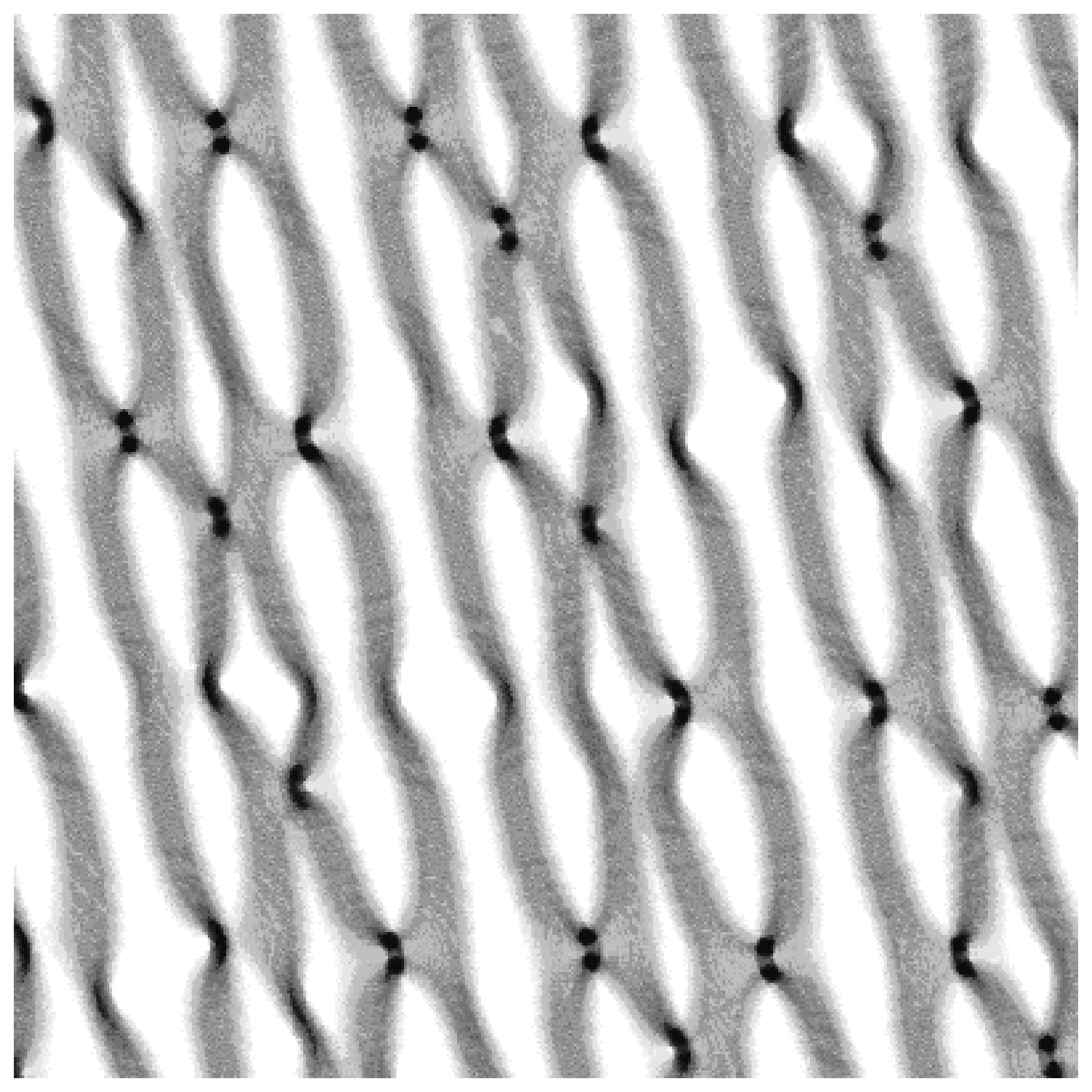} \\[3pt]

    5 3 0 &
      \includegraphics[height=2.3cm]{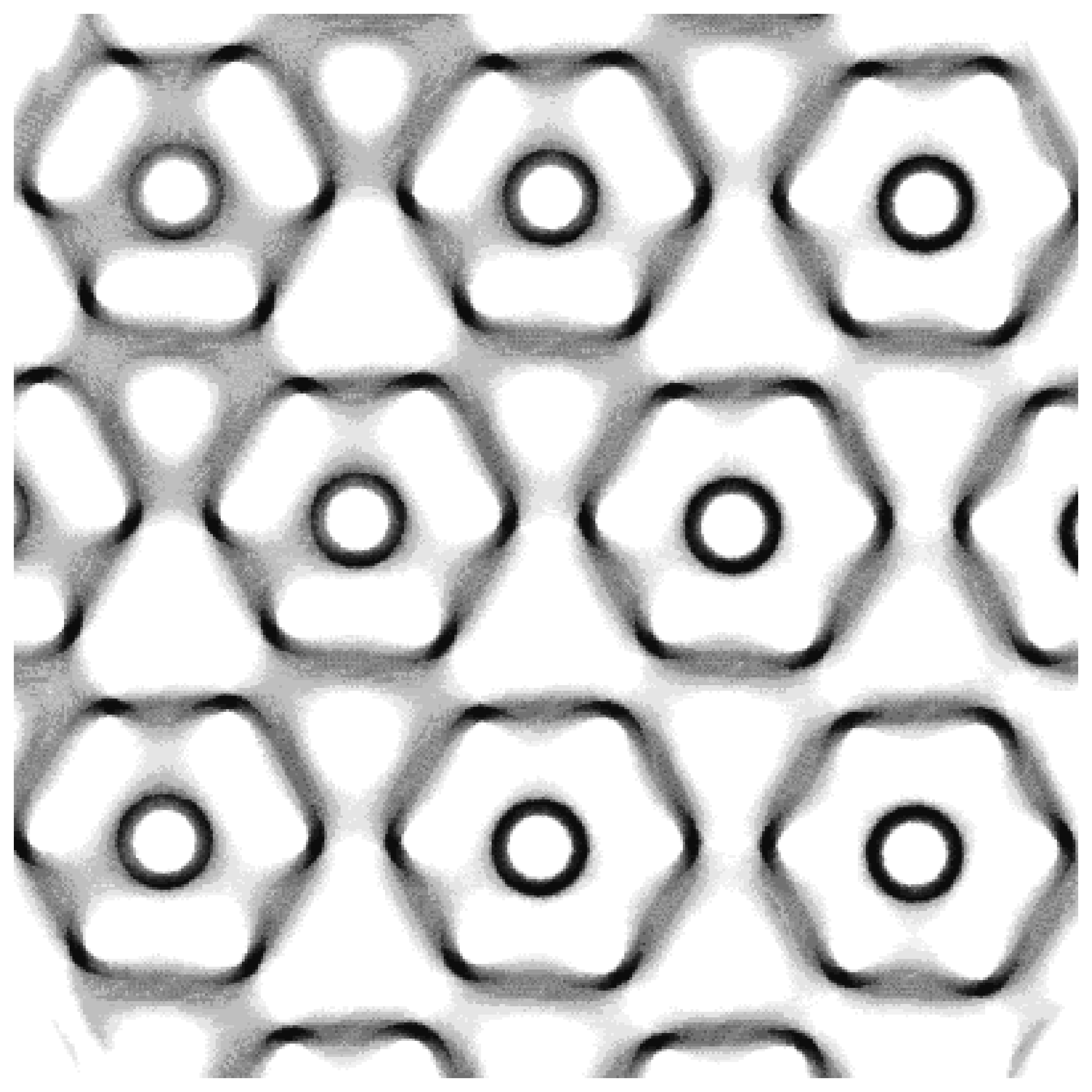} &
       \includegraphics[height=2.3cm]{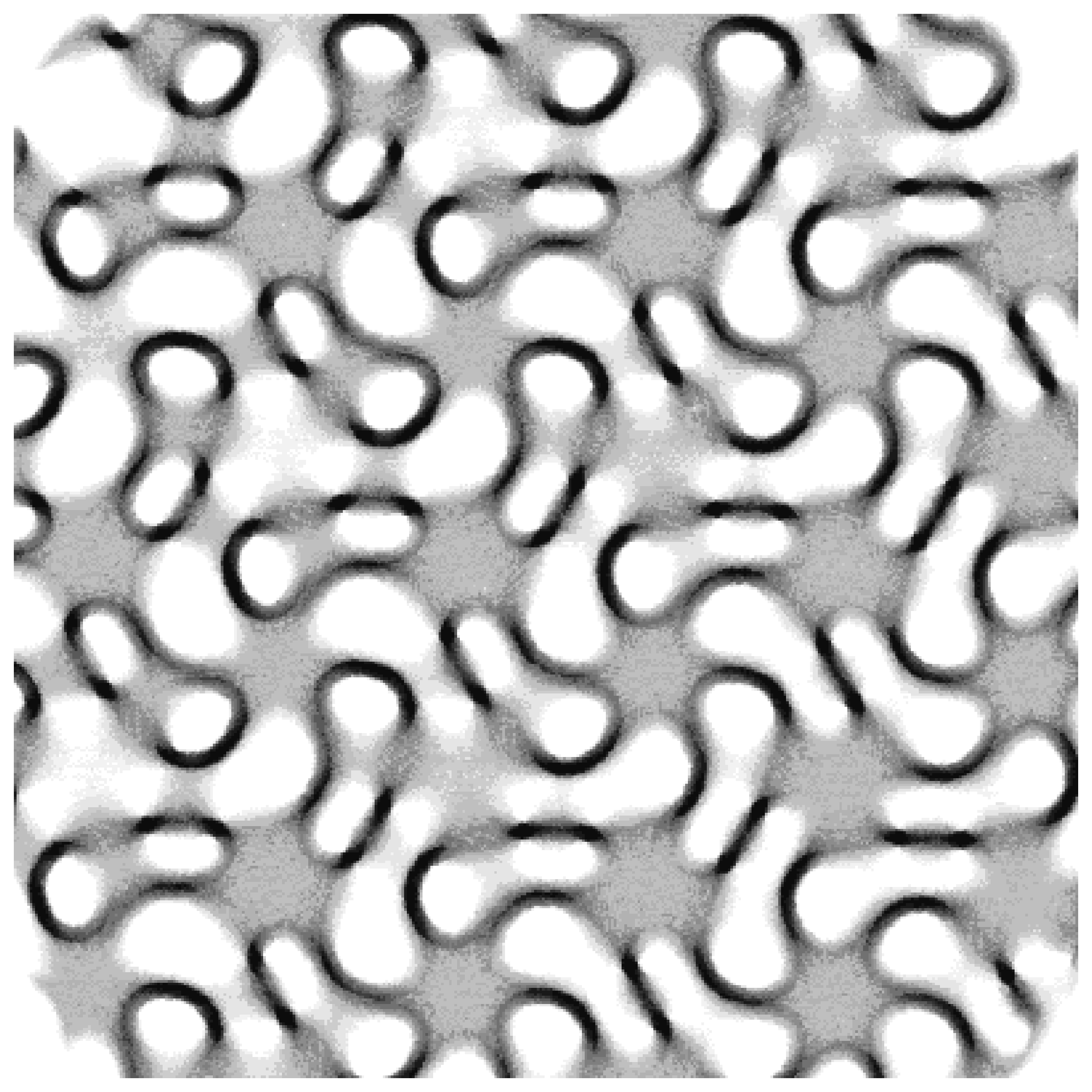} &
      \includegraphics[height=2.3cm]{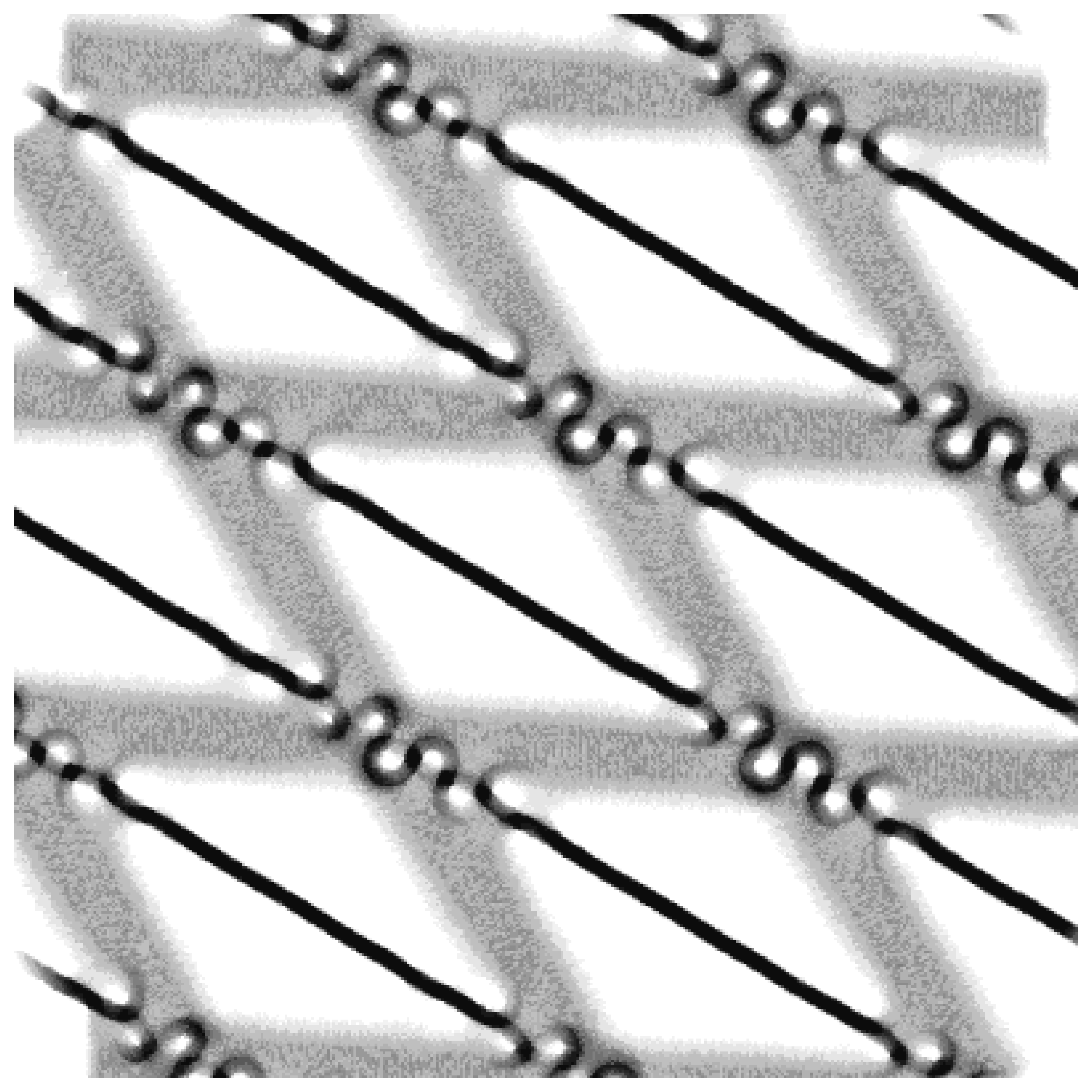} &
       \includegraphics[height=2.3cm]{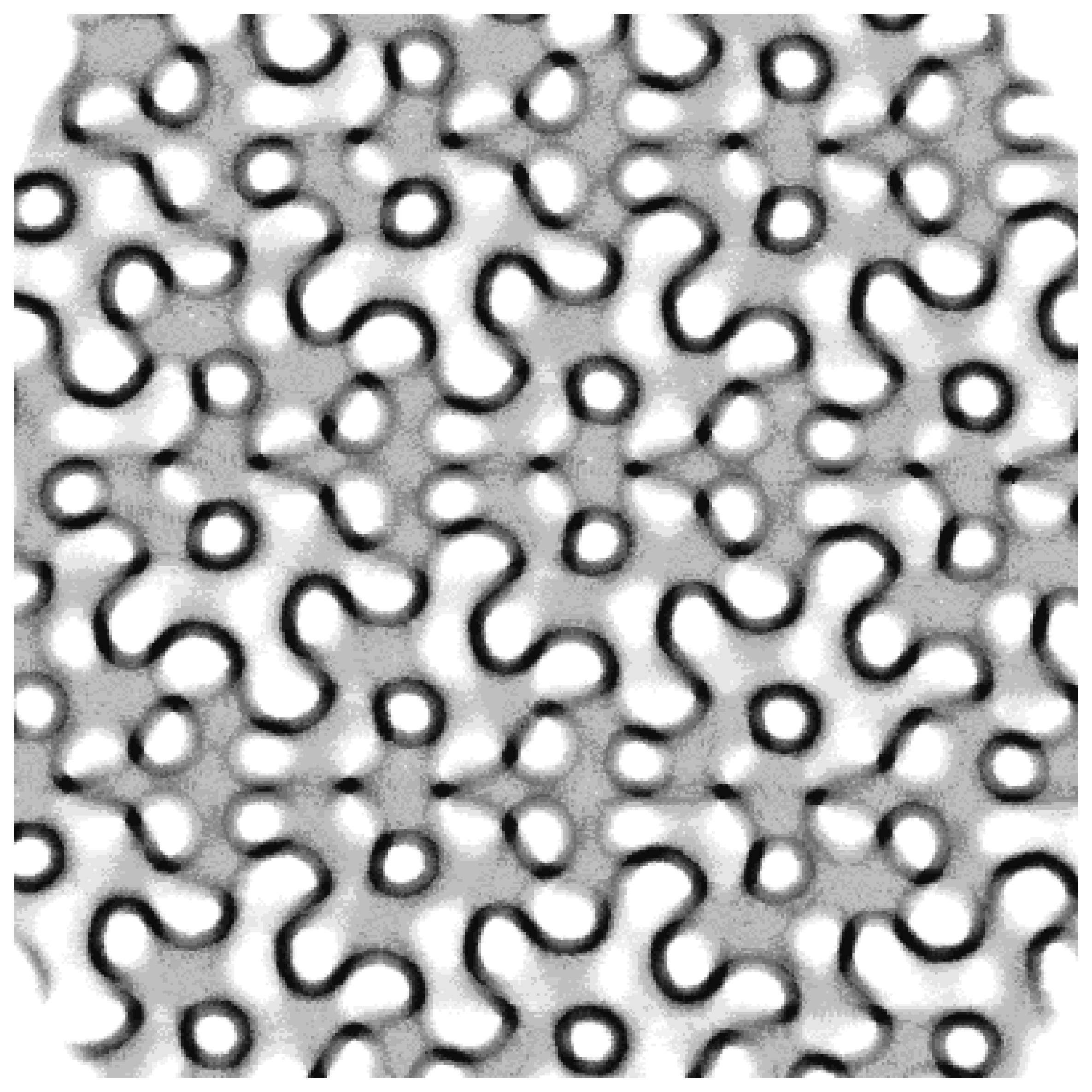}  &
      \includegraphics[height=2.3cm]{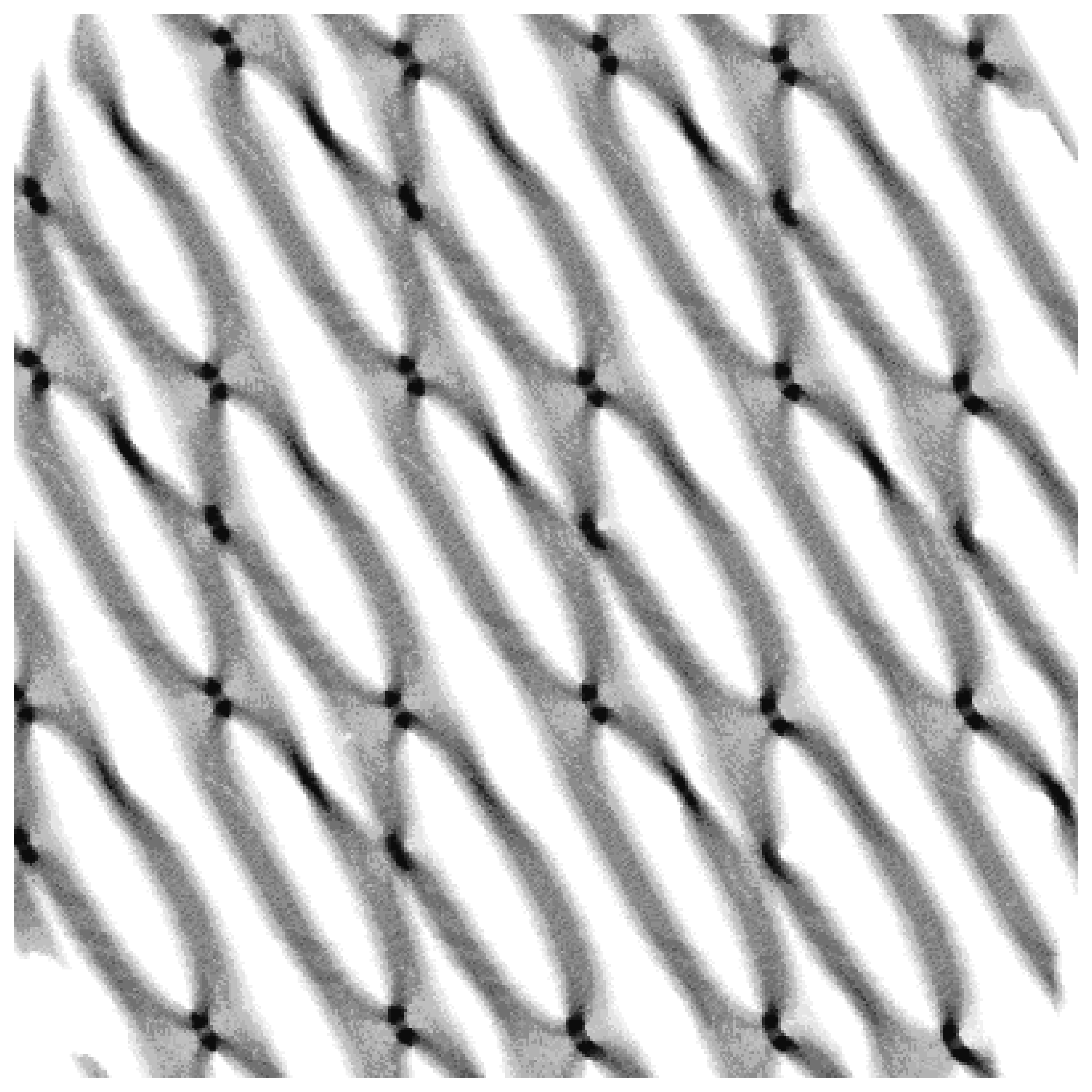}  \\
  \end{tabular}
  \caption{Gallery of selected (hkl) projections of the modelled minimal surfaces. Each triple (hkl) denotes the Miller indices that set the slice orientation. The 2‑D projections mimic transmission electron microscopy (TEM) images for easier comparison with experimental data. The first row displays the unit cell of each model, with each surface thickened to represent the membrane. All subsequent projections show the pattern produced by four unit cells; each slice is $400 \times 400 \times 70$ nm. Membrane thickness was set to 6 nm. The scale is one pixel per nanometer.}
  \label{Tab:TEM}
\end{table}

\section{Conclusions}

We explored a geometric approach that begins with the five invariant cubic rod packings (\(\Pi^*, \Pi^+, \Sigma^+, \Omega^+, \Gamma\)), inflates them into surfaces that envelope linked rod packings, and then uses \textsc{Surface Evolver} to relax them to minimal surfaces under periodic boundary conditions.  Without altering topology, each packing relaxed reproducibly to a known TPMS:
\[
  \Pi^* \;\longrightarrow\; \mathrm{I\!-\!WP}, 
  \quad
  \Pi^+,\,\Sigma^+ \;\longrightarrow\; C(^\pm Y), 
  \quad
  \Omega^+ \;\longrightarrow\; \text{H–surface}, 
  \quad
  \Gamma \;\longrightarrow\; C\bigl(I_2\!-\!Y^{\ast\ast}\bigr).
\]

These results reinforce the longstanding recognition of the intimate relationship between tubular packings and minimal surfaces. Through modeling transitions from both purely tubular structures and composite sheet–tube configurations, we observe that TPMS exhibiting cubic symmetry arise exclusively from cubic rod packings, rather than from mixed cylinder–sheet arrangements. This finding underscores the significance of investigating tubular formations during the early stages of PLB, as such studies may yield critical insights into the mechanisms underlying minimal-surface formation in membranes. Moreover, extending these analyses to rod packings with alternative symmetries, or indeed the energetic considerations of packing curvilinear cylinders \cite{evans2015ideal}, has the potential to further advance our understanding of these fundamental structural principles.

This study is subject to certain computational limitations. Modelling minimal surfaces in \textsc{Surface Evolver} inherently depends on discretisation, and small variations in the choice of unit cell can either improve or hinder convergence, although the global minimum remains unchanged. While the objective is to identify true minima with $H = 0$, in practice we accept solutions that approximate this condition sufficiently closely. Certain surfaces, such as $\Pi^{+}$ and $\Sigma^{+}$, do not converge to perfectly stable minimizers; accordingly, we select configurations that exhibit the lowest area or energy while remaining consistent with known surface geometries.

PLB often forms diamond-like structures, yet many variants remain unidentified (e.g. see \cite{gunning2001membrane}). Moreover, such non-diamond arrangements of bicontinuous membranes in plastids occur transiently both before the fully developed PLB is present (as shown above) and during the PLB-to-chloroplast thylakoid transformation upon light exposure (\cite{kowalewska2016three}. These are currently described as sponge-like configurations; however, they may represent particular types of minimal surfaces known in differential geometry but not yet recognised in biological samples. Our framework proposes additional candidates that may serve as shape descriptors of fully formed membranes. The accompanying computational approach enriches the geometric picture of possible formation pathways in biological systems.

\enlargethispage{20pt}

\section*{Declarations}

\AI{We have not used AI-assisted technologies in creating this article.}

\dataccess{Supplementary material is available online.}

\aucontribute{All authors were involved in the conceptualisation and design of the project. A.B., M.B. and Ł.K. performed the experiments, V.R. and M.E.E. performed the computational analysis, and all authors were involved in the interpretation of the results and writing the manuscript.}

\competing{We have no conflicts of interest.}

\funding{This project was funded by the National Science Centre, Poland (2022/47/I/NZ3/00498) and German Research Foundation (524578210) under the OPUS call in the Weave programme;  this article is based upon work from COST Action EuroCurvoBioNet CA22513, supported by COST (European Cooperation in Science and Technology)}

\ack{We would like to thank Matthias Himmelmann for their valuable contributions to discussions, insightful comments, and assistance with Surface Evolver. We acknowledge Houdini SideFX. TEM studies were performed in the Laboratory of Electron Microscopy of the Nencki Institute, supported by the project financed by the Minister of Education and Science based on contract no. 2022/WK/05 (Polish Euro-BioImaging Node “Advanced Light Microscopy Node Poland”).}


\printbibliography
\end{document}